\documentclass[pdflatex,sn-mathphys-num]{sn-jnl}

\usepackage{graphicx}%
\usepackage{multirow}%
\usepackage{amsmath,amssymb,amsfonts}%
\usepackage{amsthm}%
\usepackage{mathrsfs}%
\usepackage[title]{appendix}%
\usepackage{xcolor}%
\usepackage{textcomp}%
\usepackage{manyfoot}%
\usepackage{booktabs}%
\usepackage{algorithm}%
\usepackage{algorithmicx}%
\usepackage{algpseudocode}%
\usepackage{listings}%
\usepackage{hyperref}
\usepackage{etoolbox}
\let\orcidlogo\undefined
\usepackage{orcidlink}
\usepackage[normalem]{ulem}
\usepackage{float}
\usepackage[percent]{overpic}

\theoremstyle{thmstyleone}%
\theoremstyle{thmstyletwo}%

\theoremstyle{thmstylethree}%

\newcommand{\ve}[1]{\boldsymbol{#1}}
\renewcommand{\vec}[1]{\boldsymbol{#1}}
\newcommand{\ar}{$\alpha$-RuCl$_3$}
\newcommand{\tss}{\color{black}}

\begin{document}

\title[Article Title]{ Dynamical magnetotropic susceptibility as a new probe of Kitaev materials and beyond }

\author*[1]{Jo\~ao C. In\'acio  \orcidlink{0009-0006-5457-3711}}%
\email{joao.carvalho-inacio@uni-wuerzburg.de}
\author[2]{J.~Schwab \orcidlink{0000-0003-3794-8631}}%
\author[3]{G.~Rakhmanova \orcidlink{0000-0001-8750-4065}}%
\author[3]{S.~Safari }%
\author[3]{V.~Zambra \orcidlink{0000-0002-8806-5719}}%
\author[3]{H.~Nasir \orcidlink{0000-0002-3959-7129}}%
\author[4]{{\tss S.~Khim} \orcidlink{0000-0002-9495-7406}}
\author[5]{S.~Paschen \orcidlink{0000-0002-3796-0713}}
\author[3]{K. A. Modic \orcidlink{0000-0001-9760-3147}}
\author[1,6]{Fakher F. Assaad \orcidlink{0000-0002-3302-9243}}%
\author*[7]{Toshihiro Sato  \orcidlink{0000-0002-0216-176X}}%
\email{t.sato@ifw-dresden.de}

\affil*[1]{Institut f\"ur Theoretische Physik und Astrophysik, Universit\"at W\"urzburg,  97074 W\"urzburg, Germany}
\affil[2]{TU Dresden, 01062 Dresden, Germany}
\affil[3]{Institute of Science and Technology Austria, 3400 Klosterneuburg, Austria}
\affil[4]{{\tss Max-Planck-Institute for Chemical Physics of Solids, 01187 Dresden, Germany}}
\affil[5]{Institute of Solid State Physics, TU Wien, Wiedner Hauptstr.\ 8-10, 1040 Vienna, Austria}
\affil[6]{W\"urzburg-Dresden Cluster of Excellence ctd.qmat, Germany}
\affil*[7]{Institute for Theoretical Solid State Physics, IFW Dresden, 01069 Dresden, Germany}

\abstract{
{\tss The magnetotropic susceptibility, $k(\omega)$, probes ultra-low-frequency uniform ($\ve{q}=0$) spin and charge fluctuations in a crystal mounted on an oscillating cantilever: its real part shifts the oscillation frequency, while its imaginary part characterises the induced damping. We derive $k(\omega)$ within linear response theory for a generic correlated-electron Hamiltonian, showing that its real part is sensitive to magnetic anisotropy and its imaginary part encodes the uniform dynamical spin susceptibility, even for spin-symmetric insulators, while in metals it reveals eddy-current damping conditions. Using auxiliary-field quantum Monte Carlo, we compute $k(\omega)$ for microscopic models of $\alpha$-RuCl$_3$, finding that the low-temperature $k(\omega=0)/T$ scaling with $B/T$ is a signature of dominant Kitaev coupling, robust to optical phonons, while the dynamical response shows local-moment-like features. We highlight applications to Kondo destruction quantum criticality, relevant to strange metallicity and unconventional superconductivity.
}
}

\maketitle

\section*{Introduction} 

Quantum spin liquids in Kitaev materials  \cite{Matsuda25} provide a paradigmatic setting for realizing fractionalized excitations and unconventional magnetic dynamics. In the exactly solvable Kitaev model \cite{Kitaev06}, the ground state hosts emergent Majorana fermions coupled to a static $\mathbb{Z}_2$ gauge field, leading to a description in terms of non-interacting quasiparticles. However, this idealized picture is fragile: the application of a magnetic field induces interactions between Majorana fermions and flux excitations, driving the system into  a strongly correlated state. As a result, identifying the degrees of freedom underlying magnetic fluctuations in experimentally relevant settings remains a central challenge.

\begin{figure}
    \centering
    \begin{overpic}[width=0.5\textwidth]{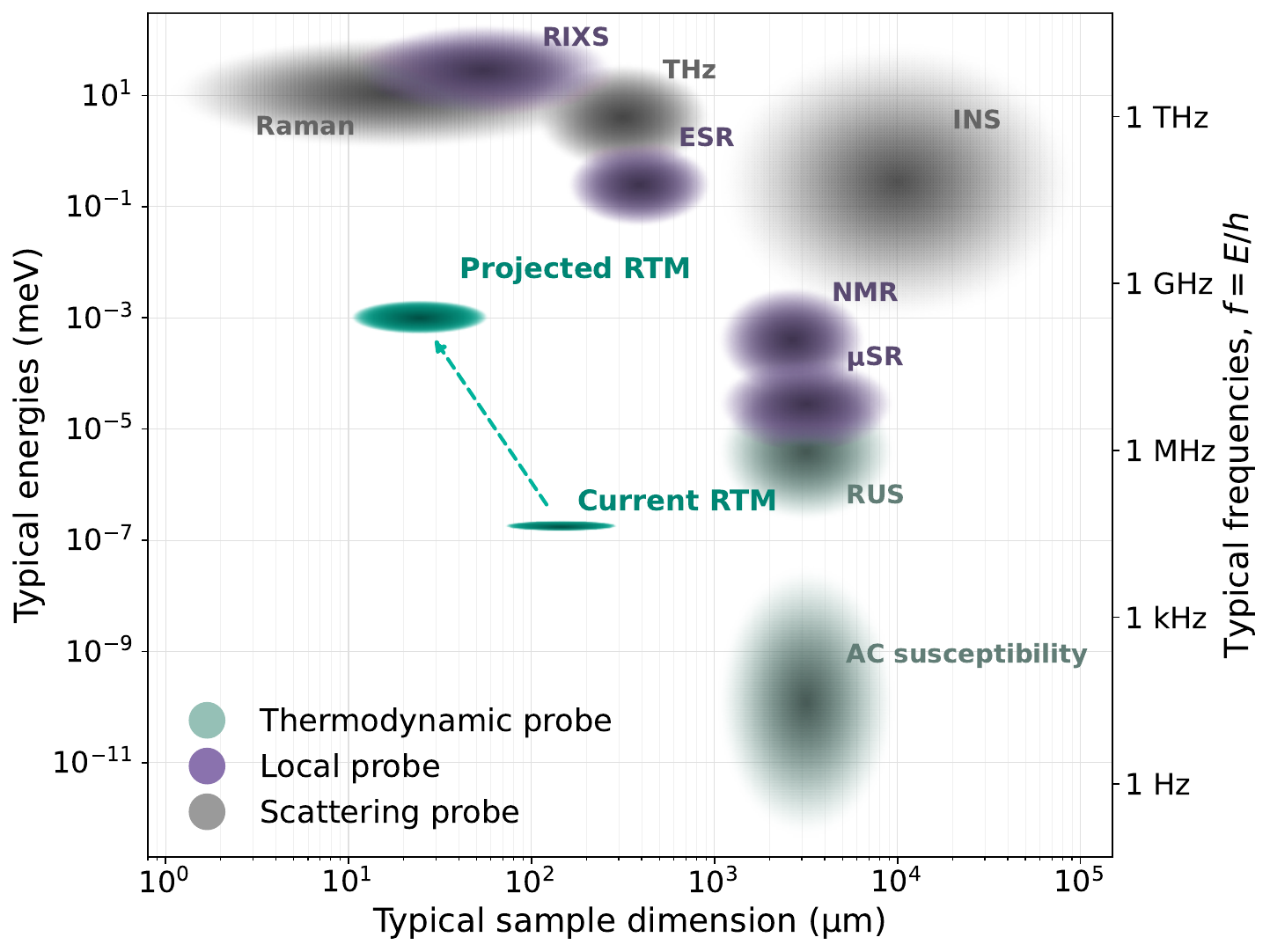}
        \put(0,71){\footnotesize \sffamily \text{(a)}}
    \end{overpic}
    \hspace{0.5cm}
    \begin{overpic}[width=0.4\textwidth]{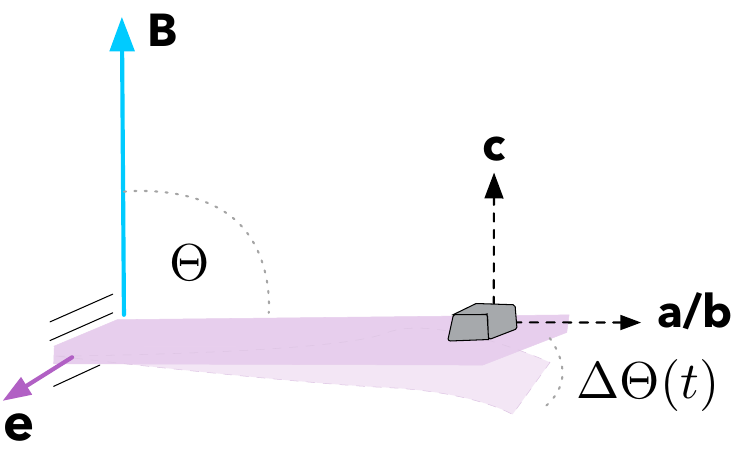}
        \put(0,89){\footnotesize \sffamily \text{(b)}}
    \end{overpic}
    \caption{
{\tss Measurement window and geometry of resonant torsion magnetometry (RTM).
(a) Landscape of dynamical probes of magnetic and lattice excitations, compared by characteristic operating frequency/energy and typical sample dimension (log-log axes; right-hand axis converts energy to frequency). Each point marks the approximate, order-of-magnitude operational range of a given technique and is colored by probe class: thermodynamic probes (teal) measure the sample's own equilibrium, uniform ($q \approx 0$) response to an applied field (AC magnetic susceptibility and Resonant Ultrasound Spectroscopy); local probes (purple) report on a specific real-space or site-resolved environment, either via an intrinsic or embedded sensor (Nuclear Magnetic Resonance and Muon Spin Resonance) or a resonant, site-selective transition (Electron Spin Resonance, Resonant Inelastic X-ray Scattering); scattering probes (gray) infer the dynamical structure factor from the energy and momentum change of an external photon or neutron (THz/far-infrared absorption, Raman scattering, Inelastic Neutron Scattering). The dynamical magnetotropic susceptibility considered in this work (detected through Resonant Torsion Magnetometry measurements) is shown at its demonstrated operating range of 40-45 kHz (Current RTM); the dashed arrow indicates the frequency range targeted by planned instrumentation development, up to ~500 MHz (Projected RTM). Values for comparison techniques are representative order-of-magnitude estimates drawn from the literature and are not intended to capture the full range of every implementation of a given method.
Resonant torsion magnetometry accesses low-frequency uniform dynamics in small crystals.
}
(b) Sketch of the cantilever setup for magnetotropic measurements. The cantilever oscillates around the axis $\ve{e}$, and the magnetic field $\ve{B}$ is applied perpendicular to this axis.}
    \label{fig:magneto}
\end{figure}

A key hurdle is the relative scarcity of experimental data on low-energy spin dynamics. While inelastic neutron scattering provides momentum-resolved information and nuclear magnetic resonance (NMR) probes local, momentum-integrated fluctuations, even less data are available for the uniform $(\vec{q}=0)$ dynamical spin correlations. The  magnetotropic susceptibility, \(k(\omega)\), offers such a probe. In particular, for insulating spin systems, the imaginary part of the dynamical response $k''(\omega)$ maps solely onto the transverse, uniform dynamical spin susceptibility, defined with respect to the direction of the magnetic field. This correspondence holds irrespective of how magnetic anisotropy is realized microscopically, establishing $k(\omega)$ as a probe of zero-momentum spin fluctuations closely related to those measured through electron spin resonance (ESR). In contrast to ESR, however, magnetotropic measurements access this response in the low-frequency limit, typically in the kilohertz range, thereby providing a window into extremely low-energy spin dynamics. 
This frequency range lies several orders of magnitude below that of conventional probes such as neutron scattering and ESR, {\tss and therefore provides a complementary window into low-frequency uniform spin dynamics that is difficult to access by other spectroscopic probes.
In Fig.~\ref{fig:magneto}(a), we provide a comparison of the operating frequency and sample size for various experimental probes of magnetic and lattice excitations. It highlights the unique operating range of resonant torsion magnetometry (RTM).}

To be concrete, the experimental setup of RTM is depicted in Fig.~\ref{fig:magneto}(b). The crystal is mounted on a cantilever that oscillates at a fixed frequency in a static magnetic field. In the rest frame of the crystal, this corresponds to a time-dependent magnetic field. Experimentally, one monitors both the shift in the oscillation frequency caused by the crystal and the damping of oscillations. Assuming small oscillation amplitudes, one can formulate a linear-response theory that relates the torque,  $\tau(\omega)$, to the oscillation amplitude, $\Delta \Theta(\omega)$,
\begin{equation}
      \tau(\omega)  =  k(\omega) \Delta \Theta(\omega).
\end{equation}
The proportionality constant $k(\omega)$ is the dynamical magnetotropic susceptibility. Its real and imaginary parts encode, respectively, the frequency shift and the damping of the oscillations. This makes the magnetotropic susceptibility a direct measurement of the low-frequency, uniform response that can be compared to microscopic model calculations.

Recent measurements of magnetotropic susceptibility in the Kitaev candidate material  $\alpha$-RuCl$_3$ have revealed unexpected behavior in precisely this low-energy regime \cite{Modic20}. Experiments and quantum Monte Carlo simulations demonstrate a striking scaling of the form $k(\omega=0)/T = f(B/T)$, which persists over a broad range of temperatures and magnetic fields~\cite{SatoT23a}. 
{\tss Below the dominant exchange scale, this scaling is reproduced in Kitaev-dominated parameter regimes, while it is not observed in the non-Kitaev benchmarks considered here.}

Despite these advances, the physical origin of this behavior lacks a clear microscopic interpretation. In particular, it remains  uncertain how the response probed by the magnetotropic susceptibility relates to  the spin dynamics in strongly correlated Kitaev systems. While the low-energy sector is expected to host fractionalized excitations, it is not clear whether and how these excitations contribute to the anisotropic response measured by $k$.

{\tss Building on the formulation of magnetotropic susceptibility by Shekhter et al.~\cite{Shekhter23}, we develop a microscopic many-body framework for the dynamical magnetotropic susceptibility $k(\omega)$ in interacting lattice models with spin and charge degrees of freedom.
In the spin-only limit, this framework clarifies the nature of the fluctuations probed by $k(\omega)$ in strongly correlated quantum magnets.}
We show that the imaginary part of the response, $k''(\omega)$, maps onto the uniform transverse spin fluctuations and that the magnetotropic response selectively emphasizes anisotropic and local contributions. 
{\tss Applying this framework to Kitaev and related models, we use $k''(\omega)$ to characterize how the uniform transverse response depends on the underlying exchange interactions, including their sign, anisotropy, and characteristic energy scales.
These results support the use of frequency-resolved magnetotropic measurements as a distinct probe for constraining candidate microscopic spin Hamiltonians and differentiating between competing interaction mechanisms.}

Beyond localized spin systems, {\tss the same framework also allows us to} analyze the magnetotropic susceptibility in representative metallic systems and compare to measurements in a clean metal, showing that the calculated response captures the measured dependence on field orientation and sample geometry. More broadly, the sensitivity of magnetotropic susceptibility to anisotropic and low-frequency fluctuations makes it a promising probe of correlated materials beyond Kitaev systems, including metals with strong spin--orbit coupling, and heavy fermion compounds. In this context, recent magnetotropic measurements in UTe$_2$ \cite{Zambra25}, which reveal transverse ferromagnetic fluctuations in the uniform channel, highlight the ability of this technique to access precisely the type of low-energy dynamics discussed here. Taken together, these results point to frequency-resolved magnetotropic measurements as a powerful tool to probe interactions and emergent correlations across a wide range of quantum materials.

The main results and organization of the article are as follows. 
{\tss We first formulate the dynamical magnetotropic response in a microscopic many-body framework for interacting lattice models,} and then focus our numerical analysis on Kitaev materials.
We provide a general definition of $k(\omega)$ for a generic Hamiltonian that encompasses the coupling of magnetic field to both charge and spin degrees of freedom. The Hamiltonian accounts for a rich variety of quantum phases, including metals, insulators, exotic phases such as strange metals, and of quantum phase transitions. We examine a range of limiting cases that establish a theoretical framework for interpreting future experiments. For spin systems, we show that the imaginary part of $k(\omega)$ maps onto the transverse uniform dynamical spin susceptibility (transverse with respect to the magnetic field). This result holds independently of the magnetic anisotropy. In the static limit, $k(\omega=0)$ vanishes for magnetically isotropic systems. For two-dimensional metallic states, the orientation of the magnetic field relative to the metal plays a dominant role, particularly through the occurrence of eddy currents. We show that distinct metallic states characterized by spin-orbit coupling and spin-nematic (i.e., altermagnetic~\cite{Libor22-2}) Fermi surfaces can exhibit distinct signatures in $k(\omega)$.

Our numerical results
{\tss use finite-temperature auxiliary-field quantum Monte Carlo (AF-QMC) simulations~\cite{Blankenbecler81,White89,Assaad08_rev,ALF_v2.4} based on a fermionic representation of spin degrees of freedom~\cite{SatoT21,Inacio25} to study microscopic spin models relevant to Kitaev materials.}
Based on numerical calculations, we argue that model systems with dominant Kitaev couplings display unique signatures in $k(\omega = 0)$. At very high temperatures, above the exchange energy scale, all quantum magnets obey the scaling law $k(\omega = 0)/T = f(B/T)$---a scaling that is expected in the paramagnetic regime. We found that models with dominant Kitaev couplings exhibit the same scaling behavior---albeit with a different scaling function $f$---also at low temperatures. We attribute this to very short-ranged magnetic correlations imposed by symmetry in the Kitaev model. 
For \ar, the  Debye temperature lies in the range of 200 K~\cite{Cao2016,Banerjee2016}, indicating phonon energy scales comparable to the dominant magnetic exchange scale. 
Therefore, coupling to phonon degrees of freedom is not expected to be negligible.
We include optical phonons in
our  quantum Monte Carlo algorithm, and explicitly show that coupling to this degree of freedom does not impair the  scaling of the 
magnetotropic susceptibility. 
Finally, we compute and discuss signatures in the imaginary part of $k(\omega)$ and confirm that when low-energy ferromagnetic spin fluctuations are present, damping becomes significant.

\section*{Results} 
\subsection*{Dynamical  magnetotropic susceptibility: definition and limiting cases}

\label{sec:magneto_def}

The  dynamical magnetotropic susceptibility  $k(\omega)$  corresponds to the  linear response of the  total torque to a temporally-varying magnetic field.  
Specifically, we  consider the setup  shown in Fig.~\ref{fig:magneto}(b) where a  crystal is  placed on a cantilever that oscillates at a given frequency $\omega$. In the presence of a magnetic field,   and in the linear-in-angle response regime, $\tau(\omega) = k(\omega) \Delta \Theta(\omega)$. Here, $\tau(\omega)$ is the Fourier transform of the total torque and $\Delta \Theta(\omega)$ is the Fourier transform of the time-dependent variation of the angle between the magnetic field and the cantilever.  The real part $k'(\omega)$ corresponds to the shift in the cantilever resonance frequency, while the imaginary part $k''(\omega)$ is related to the energy absorption by the  system 
from the oscillating cantilever, which results in damping of the cantilever oscillations. 

Classically  the cantilever corresponds to a harmonic oscillator in angular coordinates: 
\begin{equation}
    H  =   \frac{P^2}{2I}  +  \frac{k}{2} (\Delta \Theta)^2 
\end{equation}
where  $P$  is the momentum conjugate to $\Delta \Theta$, $I$ is the moment of inertia of the cantilever and the magnetotropic  susceptibility  $k$  is  given by $k=  \frac{\partial^2 H}{\partial (\Delta \Theta)^2}$.

In previous articles \cite{Shekhter23,SatoT23a}, $k(\omega)$  was discussed for quantum  spin systems.   
Here, we consider models with both spin and charge degrees of freedom to provide a broad {\tss microscopic} framework for $k(\omega)$. 
For this purpose, we consider a very generic model Hamiltonian  that can  describe both spin and  charge degrees of freedom: 
\begin{equation}
      \hat{H}   = \sum_{\ve{i},\ve{j}} \hat{\ve{c}}^\dagger_{\ve{i}} h_{\ve{i},\ve{j}} \hat{\ve{c}}_{\ve{j}} + \hat{H}_{\text{Int}}.
\end{equation}  
Here $\ve{i}$ runs over the unit cells  of a  Bravais lattice, and we will  allow $\hat{\ve{c}}^\dagger_{\ve{i}}$  to   describe an $N$-component spinor  of 
fermion  creation operators. This allows us to account for spin and orbital degrees of freedom.  In this notation, $h_{\ve{i},\ve{j}} $  is an $N \times N $  hopping matrix. To uniquely define the coupling to the electromagnetic  field,  we will require that $\hat{H}_{\text{Int}}$ enjoys local U(1) gauge invariance.  That is, $\hat{H}_{\text{Int}}$  commutes with  $\hat{T}_{U(1)} = e^{i\sum_{\ve{i},n} \varphi_{\ve{i},n} \hat{c}^{\dagger}_{\ve{i},n} \hat{c}^{\phantom{\dagger}}_{\ve{i},n  }}$ where $n$ runs from $1\cdots N$ and $\varphi_{\ve{i},n}$ is an arbitrary unit cell and orbital-dependent phase.  This constraint is satisfied by a large class of interaction terms, including density-density interactions,   any  spin Hamiltonian,  as well as a class of  electron-phonon and  spin-phonon couplings. Our  derivation is based on standard Kubo linear response theory \cite{Mahan90}.

Owing to the oscillation of the cantilever, the magnetic field perceived by  the system is  time  dependent and can be expressed as: 
\begin{equation}
    \ve{B}(t) = e^{i \ve{K}  \cdot \ve{e} \Delta \Theta(t)  } \ve{B},
\end{equation}
where $\ve{K} = \left( K^{1}, K^{2}, K^{3} \right)$ is a  vector of the generators of SO(3) rotations, $K^{n}_{m,m'} = -i\epsilon_{n,m,m'} $ with $\epsilon_{n,m,m'}$ the  antisymmetric tensor, $\ve{e}$ the axis of rotation, and $\Delta \Theta(t) $  the time-dependent variation of the angle due to the oscillations of the cantilever. 
As  depicted in the sketch of Fig.~\ref{fig:magneto}(b), we will consider the case where $\ve{e} \cdot \ve{B} = 0$.  This choice is not necessary, but it simplifies the expressions for the magnetotropic susceptibility. {\tss We note that our calculations are carried out in the oscillating frame of the crystal. In this frame the vector potential becomes time dependent thereby generating an effective electric field. Furthermore, since this frame is not an inertial system, one should take into account fictitious forces. In the Supplementary Material, we show that these effects are negligible even for the projected  RTM  measurements of Fig.~\ref{fig:magneto}(a).}

To account for the  orbital  coupling, we  have to derive the time-dependent vector  potential $\ve{A}(\ve{x},t)$ that produces the magnetic field $\ve{B}(t)$.  Let  $\ve{B}(t) =  \ve{\nabla} \times \ve{A}(\ve{x}, t) $, then the curl of the time-dependent vector potential  
\begin{equation}
  \label{Eq:vec_pot_rot_frame}
   \ve{A}(\ve{x},t) =  R(t)  \ve{A}\left(R^{T}(t)  \ve{x} \right) 
\end{equation} 
will produce $\ve{B}(t)$. The   coupling to the magnetic  field includes  the  Peierls substitution as well as the  Zeeman term 
\begin{equation}
   \hat{H} (t)  = \sum_{\ve{i},\ve{j}} \hat{\ve{c}}^\dagger_{\ve{i}} h_{\ve{i},\ve{j}}   e^{\frac{2 \pi i}{\Phi_0} \int_{\ve{i}}^{\ve{j}} \ve{A}(\ve{x},t) \cdot d\ve{x}}\hat{\ve{c}}_{\ve{j}} + \hat{H}_{\text{Int}}  + \mu_B g \ve{\hat{S}}_{\text{tot}} \cdot \ve{B}(t).
\end{equation}
In the above,  $\Phi_0$ corresponds to the flux quantum, $\ve{\hat{S}}_{\text{tot}} $ is the total spin operator, $g$  the  gyromagnetic tensor, and $\mu_B$  the Bohr magneton.

We will consider small oscillations such that $\Delta \Theta(t) \ll 1$ and  expand  in this quantity. 
To proceed  with the calculation, it is convenient to choose a gauge  in which  $\ve{e} \times \ve{A}(\ve{x}) = 0$. Note that $\ve{\nabla} \times \ve{A}(\ve{x}) = \ve{B}$.  In this gauge, an expansion in $\Delta \Theta(t)$ up to second order gives
\begin{eqnarray}
        \ve{A}(\ve{x},t) &=& \ve{A}(\ve{x}) + \Delta \Theta(t)  \frac{\partial\ve{A}}{\partial x_n} (\ve{e}\times \ve{x})_n +  \\
        & &  \frac{\Delta \Theta^2(t)}{2} \left(\frac{\partial\ve{A}}{\partial x_n} ( \ve{e}\times (\ve{e}\times \ve{x}))_n  +  \frac{\partial^2\ve{A}}{\partial x_n \partial x_m} (\ve{e}\times \ve{x})_n (\ve{e}\times \ve{x})_m  \right) + {\cal O} \left(\Delta \Theta^3 \right) \nonumber.
\end{eqnarray}
In the above, the sum over repeated spatial indices is implied.    The perturbation to the Hamiltonian is  given  by:
\begin{multline}
 \hat{H}(t) =  \underbrace{\sum_{\ve{i},\ve{j}} \hat{\ve{c}}^\dagger_{\ve{i}} h_{\ve{i},\ve{j}}   e^{\frac{2 \pi i}{\Phi_0} \int_{\ve{i}}^{\ve{j}} \ve{A}(\ve{x}) \cdot d\ve{x}}\hat{\ve{c}}_{\ve{j}} + \hat{H}_{\text{Int}}  + \mu_B g \ve{\hat{S}}_{\text{tot}} \cdot \ve{B}}_{\equiv \hat{H}_{0}}  + \\
 \underbrace{\hat{\tau}_P \Delta \Theta(t) }_{\equiv \hat{H}_1} + \hat{\tau}_D \frac{\Delta \Theta(t)^2}{2} + {\cal O} \left( \Delta \Theta^3\right)
\end{multline}
with 
\begin{equation}
    \hat{\tau}_P =  - \mu_B g \ve{\hat{S}}_{\text{tot}} \cdot \left(  \ve{e} \times \ve{B} \right) +  
    \frac{2 \pi i}{\Phi_0} \sum_{\ve{i},\ve{j}} \hat{\ve{c}}^\dagger_{\ve{i}} \left( h_{\ve{i},\ve{j}}    e^{\frac{2 \pi i}{\Phi_0} \int_{\ve{i}}^{\ve{j}}  \ve{A}(\ve{x}) \cdot d\ve{x}}    
    \int_{\ve{i}}^{\ve{j}}\frac{\partial\ve{A}(\ve{x})}{\partial x_n} ( \ve{e}\times \ve{x})_n 
    \cdot d \ve{x} \right) 
      \hat{\ve{c}}_{\ve{j}}
\end{equation}
and
\begin{eqnarray}
 & & \hat{\tau}_D = \mu_B g \ve{\hat{S}}_{\text{tot}} \cdot \ve{e} \times \left( \ve{e} \times \ve{B}  \right)   
+   \frac{2 \pi i}{\Phi_0} \sum_{\ve{i},\ve{j}} \hat{\ve{c}}^\dagger_{\ve{i}} h_{\ve{i},\ve{j}}  \left(  e^{\frac{2 \pi i}{\Phi_0} \int_{\ve{i}}^{\ve{j}}  \ve{A}(\ve{x}) \cdot d\ve{x}} \right)  \times 
     \nonumber  \\  && \left[ \frac{2 \pi i}{\Phi_0} \left( \int_{\ve{i}}^{\ve{j}}\frac{\partial\ve{A}(\ve{x})}{\partial x_n} ( \ve{e}\times \ve{x})_n 
    \cdot d \ve{x} \right)^2  + \int_{\ve{i}}^{\ve{j}} \frac{\partial^2\ve{A}(\ve{x})}{\partial x_n \partial x_m} (\ve{e}\times \ve{x})_n (\ve{e}\times \ve{x})_m  \cdot d\ve{x}  \right. \nonumber  \\ 
    & & \left. + \int_{\ve{i}}^{\ve{j}}  
   \frac{\partial\ve{A}}{\partial x_n} ( \ve{e}\times (\ve{e}\times \ve{x}))_n   \cdot d \ve{x} \right] \,
      \hat{\ve{c}}_{\ve{j}}.  \nonumber 
\end{eqnarray}
Here we  have  adopted the notation of the optical conductivity \cite{Scalapino93} and  introduce the paramagnetic $\hat{\tau}_P$ and diamagnetic $\hat{\tau}_D$ torque operators.  
 The total torque operator reads
 \begin{equation}
  \hat{\tau}   =  \frac{\partial \hat{H}}{\partial \Delta \Theta(t)} =   \hat{\tau}_P + \Delta \Theta(t) \hat{\tau}_D  + {\cal O}  \left( \Delta \Theta^2\right) .
 \end{equation}

Consider an oscillation  with a single Fourier component  $\Delta \Theta(t) = \Delta \Theta  e^{i \omega t}$.
The  Kubo linear response of the total torque density  to the oscillations of the cantilever  now  reads
\begin{equation}
    \frac{1}{V}\langle \hat{\tau} \rangle (t)  = \frac{1}{V} \langle \hat{\tau}_P \rangle_0 + k(\omega)e^{i \omega t} \Delta \Theta 
\end{equation} 
with the magnetotropic dynamical susceptibility
\begin{equation}
    k(\omega) = \frac{1}{V}  \left[ \langle \hat{\tau}_D \rangle_0 - i \int_0^{\infty} dt' \, e^{-i \omega t'} \langle [ \hat{\tau}_P, \hat{\tau}_P(-t') ] \rangle_0 \right].
\label{eq:kom}
\end{equation}

If  the frequency of the cantilever oscillations is small compared to the characteristic frequencies of the system, we can consider the limit 
of time-independent magnetic fields. In this case, the  uniform  magnetotropic susceptibility  is given by
\begin{equation}
    k  = \frac{1}{V}\left. \frac{ \partial^2 F}{\partial \Delta \Theta^2}  \right|_{\Delta \Theta = 0}  = 
    \frac{1}{V}\left[ \langle  \hat{\tau}_D \rangle_0 -  \int_0^{\beta} d\tau  \langle \Delta \hat{\tau}_{P}(\tau) \Delta \hat{\tau}_{P} \rangle_0  \right],
\end{equation}
where $F$ is the free energy of the system  and $\Delta \hat{\tau}_{P}(\tau) = e^{\tau \hat{H}_0} \left( \hat{\tau}_{P} - \langle \hat{\tau}_{P} \rangle_0\right) e^{-\tau \hat{H}_0}$.   Using the Lehmann representation, $\hat{H}_0 | n \rangle = E_n | n \rangle$,  
one can show that 
\begin{eqnarray}
   \lim_{\omega \to 0} i \int_0^{\infty} dt' \, e^{-i \omega t'} \langle [ \hat{\tau}_{P}, \hat{\tau}_{P}(-t') ] \rangle_0  & &  = 
  \lim_{\omega \to 0}  
   \frac{1}{Z} \sum_{n,m} |\langle n | \Delta \hat{\tau}_{P} | m \rangle|^2     \frac{ e^{-\beta E_n}  - e^{-\beta E_m}   }{  E_m - E_n + \omega - i 0^+}  \nonumber \\
   & & =
   \frac{1}{Z} \sum_{n \neq m} |\langle n | \Delta \hat{\tau}_{P} | m \rangle|^2     \frac{ e^{-\beta E_n}  - e^{-\beta E_m}   }{  E_m - E_n }
\end{eqnarray}
and that
\begin{eqnarray}
    \int_0^{\beta} d\tau   \langle \Delta \hat{\tau}_{P}(\tau) \Delta \hat{\tau}_{P} \rangle_0 & & = 
   \frac{1}{Z} \sum_{n \neq m} |\langle n | \Delta \hat{\tau}_{P} | m \rangle|^2  \,    \frac{ e^{-\beta E_n}  - e^{-\beta E_m}   }{  E_m - E_n } \nonumber \\
   & & + \frac{\beta}{Z} \sum_{n} e^{-\beta E_n} | \langle n | \Delta \hat{\tau}_{P} | n \rangle |^2.
\end{eqnarray}
It is known that response functions can be singular in the zero-frequency uniform limit \cite{Scalapino93}, and the above equation is yet another 
manifestation of this fact. 

In this article,  we will compute both $k$  and  $k''(\omega) = \text{Im} [k(\omega)]$  for various models of quantum magnets and  metals. 
To compute $k''(\omega)$, we will use the Algorithms for Lattice Fermions \cite{ALF_v2.4} implementation of the Stochastic Analytical Continuation (SAC) method \cite{Beach04a,Sandvik98,Shao23}, which estimates the imaginary part $k''(\omega)$ given the imaginary-time fluctuations
\begin{equation}
    \langle \Delta \hat{\tau}_{P}(\tau) \Delta \hat{\tau}_{P} \rangle_0  = \int_{-\infty}^{\infty} d \omega \, k''(\omega) \, K(\omega, \tau)
\end{equation}
with  Kernel $K(\omega, \tau) = \frac{1}{\pi} \frac{e^{-\tau \omega}}{1 - e^{-\beta \omega}}$

For  generic spin models, the  relation of  the dynamical magnetotropic susceptibility to the dynamical spin susceptibility  is given by 
\begin{equation}
    k(\omega) = \frac{1}{V}  \left[ \mu_B g \langle \ve{\hat{S}}_{\text{tot}} \rangle_0 \cdot \ve{e} \times \left( \ve{e} \times \ve{B}  \right)  + \mu_B^2  (\ve{e}\times \ve{B} )^T  g \chi(\omega) g^{T} ( \ve{e}\times \ve{B} ) \right] 
\label{eq:imag_k}    
\end{equation}
with  the  dynamical spin susceptibility tensor defined as
\begin{equation}
    \chi^{\alpha \beta}(\omega) = - i \int_0^{\infty} dt' \, e^{-i \omega t'} \langle [ \hat{S}^{\alpha}_{\text{tot}}, \hat{S}^{\beta}_{\text{tot}}(-t') ] \rangle_0.
\end{equation}
The  above  reveals  that $k(\omega)$  probes $\ve{q}=0$ spin fluctuations \textit{transverse} to the  magnetic field direction or, more precisely, in the direction of   $g^{T}(\ve{e} \times \ve{B})$ for an anisotropic  $g$-tensor.

The uniform magnetotropic susceptibility is sensitive to magnetic anisotropy. Consider an SU(2)-symmetric spin system $\hat{H}_{\text{Spin}}$. Under this assumption, the partition function $Z(g\ve{B})$ of the system in a magnetic field satisfies the relation $Z(g\ve{B}) = Z( R g\ve{B}) $ with R an arbitrary rotation matrix.  Furthermore, if $R g  R^{T} = g$, then 
$Z(g\ve{B}) = Z(g R \ve{B})$.  This implies that the free energy $F = - \beta^{-1} \ln Z$  only depends on the magnitude of the magnetic field $B = |\ve{B}|$.  As a consequence, the magnetotropic susceptibility $k$ vanishes identically for SU(2)-symmetric spin systems.  

Let  us now consider  a local  moment, with 
\begin{equation}
  \hat{H}_{\text{Spin}}(t) = - g \ve{\hat{S}} \cdot \ve{B}(t).
\end{equation}
In the above, we have absorbed $g \to \mu_B g$, for simplicity. 
For this local moment Hamiltonian  and $\ve{B}(t)  =  e^{i \ve{K} \ve{e} \Delta \Theta (t) } \ve{B}  $  the dynamical 
magnetotropic susceptibility reads 
\begin{equation}
k(\omega) =  g \langle \ve{\hat{S}} \rangle_0 \cdot \ve{e} \times \left( \ve{e} \times \ve{B}  \right) 
+ (\ve{e}\times \ve{B})^{T} g  \frac{1}{\omega +  g^{T}\ve{B}  \cdot  \ve{K} - i 0^{+}} \left(  g^{T}( \ve{e}\times \ve{B}) \cdot \ve{K} \right) \langle \ve{\hat{S}} \rangle_0.
\end{equation}
It is interesting to  consider  a  specific  example with $g = \text{diag}(g_{1}, g_{2}, g_{3})$, $\ve{B} = B \ve{e}_3$, and $\ve{e} = \ve{e}_1$. 
In this case, we obtain
\begin{equation}
  k(\omega)  = \frac{\tanh\left( \frac{\beta  g_3 B}{2} \right) }{2} 
    \left[ - g_3  B  +   \frac{(g_2  B)^2}{2} \left(  \frac{1}{\omega + g_3  B + i 0^{+}}  - \frac{1}{\omega - g_3   B + i 0^{+}}  \right)  \right].  
\label{eq:local_moment_k}
\end{equation}
From this we can conclude that i) the magnetotropic susceptibility $k(\omega = 0) $  takes a finite value  only in the anisotropic case $g_2 \neq g_3$; ii) the imaginary part $k''(\omega)$  displays two peaks at frequencies $\omega = \pm g_3  B$ corresponding to the Larmor precession of the local moment around the magnetic field; iii) $k''(\omega)$  is finite even  in the isotropic case; and   iv)  the scaling relation $\beta k = f(\beta B)$  holds for $k(\omega=0)$. Note that such a scaling  relation is a signature of the locality of spin fluctuations, not  a signature of scale invariance generically  observed at critical points.

{\tss
\subsection*{Spin liquids: parton  mean-field approximation}

\begin{figure}
\centering
\includegraphics[width=0.5\linewidth]{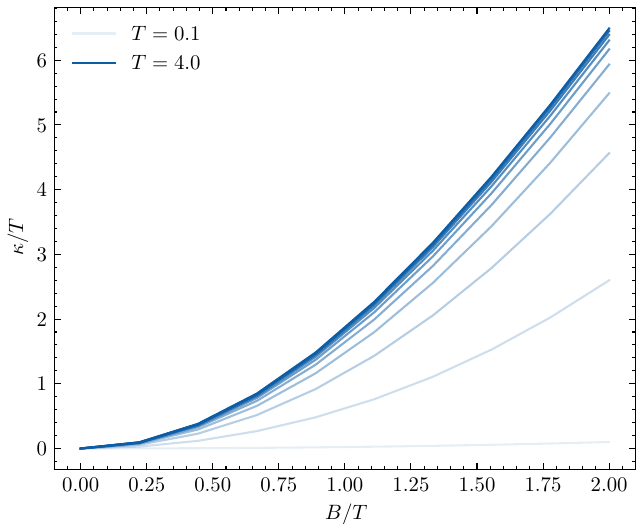}
\caption{\label{fig:spin_liquid}
{\tss
Magnetotropic susceptibility of a U(1) Dirac spin liquid on the honeycomb lattice in the parton mean-field approximation. The  calculations are carried out on an L=24  lattice  with $\ve{e} = \ve{e}_a$, and we have set $J\chi = 1$.  } }
\end{figure}

Next we consider a parton mean-field approximation of the spin-liquid state along the lines of Ref.~\cite{Marston89}.  Our starting  point is the Heisenberg Hamiltonian on the honeycomb lattice 
\begin{equation}
   \hat{H} = J \sum_{\langle \ve{i},\ve{j} \rangle} \ve{\hat{S}}_{\ve{i}} \cdot \ve{\hat{S}}_{\ve{j}} -  \mu_B \sum_{\ve{i}} \ve{\hat{S}}_{\ve{i}} \cdot \ve{g}  \ve{B}
\end{equation}
with $\langle \ve{i},\ve{j} \rangle$ denoting nearest-neighbor pairs and  
$\ve{g} = \text{diag}(2.3, 2.3, 1.3) $ the $g$-tensor  and  $\ve{B} = B \ve{e}$.   For the lattice model  in the x-y  plane, we  require  $\ve{e} $  not  to be orthogonal to this plane to ensure  that the magnetotropic susceptibility does not vanish by symmetry.   Specifically, we have chosen  the  direction  considered in  Fig.~2a of Ref.~\cite{SatoT23a}, corresponding to $\ve{e}^T =  \ve{e}_a^T = \frac{1}{\sqrt{6}}\left( 1,1,-2\right)$.  
Adopting the Abrikosov fermion representation of the spin operators, $\ve{\hat{S}}_{\ve{i}} = \frac{1}{2} \ve{\hat{f}}^\dagger_{\ve{i}} \ve{\sigma} \ve{\hat{f}}_{\ve{i}}$, in the  Hilbert space  with one  fermion per  site, we obtain the following mean-field Hamiltonian:
\begin{equation}
   \hat{H}_{\text{spinon}} = J \chi \sum_{\langle \ve{i},\ve{j} \rangle} \ve{\hat{f}}^\dagger_{\ve{i}} \ve{\hat{f}}_{\ve{j}} + \text{h.c.}  
   -  \mu_B \sum_{\ve{i}} \ve{\hat{S}}_{\ve{i}} \cdot \ve{g}  \ve{B}
\end{equation}
with the  constraint $\frac{1}{N}\sum_{\ve{i}} \langle \ve{\hat{f}}^\dagger_{\ve{i}} \ve{\hat{f}}_{\ve{i}} \rangle= 1$  imposed on average.  This mean-field Hamiltonian, describing free spinons, corresponds to the mean-field approximation of a so-called U(1) Dirac spin liquid \cite{Xu18}  where the spinons are  described by an emergent  Dirac Hamiltonian at low energies.    Within this approximation, we can compute the 
magnetotropic susceptibility.  While our results, see  Fig.~\ref{fig:spin_liquid},  show the usual high-temperature scaling,
it is absent at low temperatures.  We conclude that at least at the mean-field level, scaling is not a signature of fractionalization and concomitant spin-liquid physics.}

\subsection*{Two-dimensional metallic systems: PdCoO$_2$}
\begin{figure}
\centering
\includegraphics[width=0.9\linewidth]{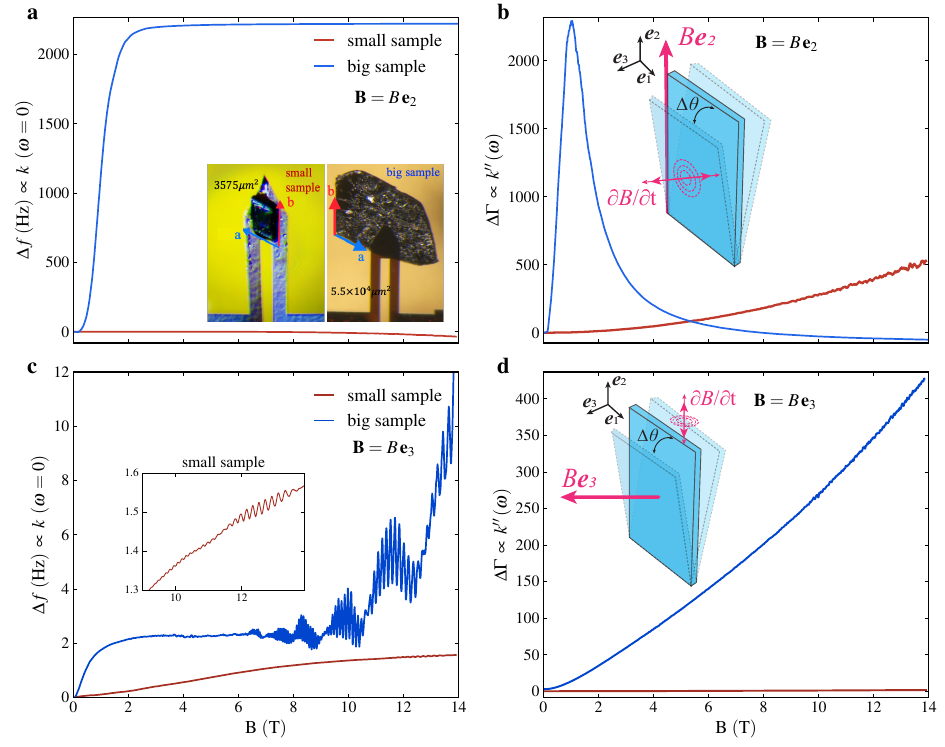}
\caption{\label{fig:Metal_PdCoO2}
Magnetic-field dependence of the magnetotropic susceptibility of the clean metal {\tss PdCoO$_2$} for in-plane and out-of-plane field orientations.
Panels \textbf{(a)} and \textbf{(b)} show the shift in the cantilever resonance frequency, $\Delta f \propto k'(\omega = 0)$, and the damping of the cantilever oscillations, $\Delta \Gamma = k''(\omega)$, for the in-plane configuration, $\mathbf{B} = B\,\mathbf{e}_2$. Panels \textbf{(c)} and \textbf{(d)} show the corresponding response for the out-of-plane configuration, $\mathbf{B} = B\,\mathbf{e}_3$. The inset in panel \textbf{(c)} highlights quantum oscillations in the small-sample signal. The panels \textbf{(b)} and \textbf{(d)} include schematic illustrations of the corresponding field orientation. Data for the big and small samples are compared in each panel.  All data are shown at $T = 1.5\,\mathrm{K}$.
Both samples have thickness, $t \approx 14\,\mu\mathrm{m}$, but differ strongly in lateral size: the big sample has area $S \approx 5.5 \times 10^4\,\mu\mathrm{m}^2$ and volume $V \approx 7.5 \times 10^5\,\mu\mathrm{m}^3$, whereas the small sample has area $S = 3575\,\mu\mathrm{m}^2$ and volume $V \approx 5 \times 10^4\,\mu\mathrm{m}^3$, corresponding to a reduction in volume by roughly a factor of 15. 
For the big and small samples, the resonance frequencies are  42550 Hz and 44046 Hz respectively. 
}
\end{figure}

Let us now consider a two-dimensional  metallic  system lying in the \(x\)-\(y\) plane    and  various   orientations of the magnetic field  both 
from the  experimental and theoretical point of view. For the experimental part, we consider the metallic delafossite  PdCoO$_2$ known for its exceptionally high in-plane conductivity, extremely long mean free path, and quasi-two-dimensional electronic structure  \cite{Hicks12,Mackenzie17}. 
It thus provides an ideal platform for quantitative comparison between theoretical predictions and experimental observations. The details of the experimental setup and sample preparation are provided in the Methods section.

Considering \(\vec{e} = \vec{e}_1\) and \(\vec{B} = B \vec{e}_2\), our  gauge  choice, $\ve{e} \times \ve{A}(\ve{x}) = 0$,   is  satisfied for $\ve{A}(\ve{x}) = B(x_3,0,0)$.  In this case, we have 
\begin{equation}
    \frac{\partial\ve{A}(\ve{x})}{\partial x_n} ( \ve{e}\times \ve{x})_n  =  B \ve{e}_1 (\ve{e}_1 \times \ve{x})_3 =   B \ve{e}_1 x_2
\end{equation}
and 
$ \frac{ \partial^2\ve{A}(\ve{x})}{\partial x_n \partial x_m} = 0$. 
Since the vector potential  depends only on $x_3$  and  the metal  lies in the \(x\)-\(y\) plane,  we can omit it from the Hamiltonian, e.g. by choosing $x_3 = 0$, such that    $\hat{H}_0$ only contains the Zeeman term.   Setting $x_3=0$ again yields: 
\begin{eqnarray}
    && \hat{\tau}_P = 
    - g \ve{\hat{S}}_{\text{tot}} \cdot \left(  \ve{e} \times \ve{B} \right) +  
    \frac{ B \pi i}{\Phi_0} \sum_{\ve{i},\ve{j}} \hat{c}^\dagger_{\ve{i}} h_{\ve{i},\ve{j}}   
     \left(\ve{j}-\ve{i}\right)_1\left(\ve{j}+\ve{i}\right)_2 \hat{c}_{\ve{j}}   \text{  and } \\
    && \hat{\tau}_D = g \ve{\hat{S}}_{\text{tot}} \cdot \ve{e} \times \left( \ve{e} \times \ve{B}  \right) 
  -  \left(\frac{B \pi}{\Phi_0}\right)^2 \sum_{\ve{i},\ve{j}} \hat{c}^\dagger_{\ve{i}} h_{\ve{i},\ve{j}}
         \left(\ve{j}-\ve{i}\right)_1^2\left(\ve{j}+\ve{i}\right)_2^2 \hat{c}_{\ve{j}}.     \nonumber 
\end{eqnarray}

For this  geometry, the variation of the magnetic flux piercing the 2D metal  is proportional to $\Delta\Theta(t)$, such that eddy currents appear in the torque operator $\hat{\tau}_P$.
It is interesting to note  that  the eddy current operator scales as $(\ve{i} + \ve{j})_2$, and  the total  torque $\hat{\tau}_D$ as  $(\ve{i} + \ve{j})_2^2$.  This has important consequences for  the scaling of the magnetotropic susceptibility. 
Since  we have  normalized by the volume in the definition of $k(\omega)$ in Eq.~\ref{eq:kom}, we expect $k(\omega)$ to   be an 
intensive  quantity. This is certainly  the case  for  spin  systems.  For  metallic  systems, however,  the eddy current contribution  to $\hat{\tau}_P$  produces  a  result  that scales 
more  quickly than the volume, such that even  after normalization  $k(\omega)$ can diverge  for  big  systems.  
Classically, eddy currents stem  from Faraday's  law of induction:  the induced voltage is proportional  to the time variation of the magnetic flux piercing the system, $\Phi(t)$.  
For our geometry, $\Phi(t) = L_1 L_2 B \sin(\Delta \Theta(t))$ with $L_1$ and $L_2$ the linear dimensions of the system.  Owing to Lenz's law, the induced current's magnetic field opposes the variation of the magnetic flux, and thereby generates a Lorentz force that damps the oscillations of the cantilever.    

Fig.~\ref{fig:Metal_PdCoO2} (a),(b) show experimental results for this geometry.  As hinted by the theory, one  observes a strong size  dependence.  For the larger 
crystal, the frequency shift $\Delta f \propto k(\omega = 0)$ as a function of the magnetic field $B$ grows  rapidly and then  saturates.   Before  saturation, damping  grows 
considerably, Fig.~\ref{fig:Metal_PdCoO2} (b), and is then suppressed at higher  fields. An interpretation of the above  is  that the big enhancement of $k(\omega = 0)$  under magnetic  field corresponds to an increasing stiffness of the cantilever as the induced current's magnetic field continues to oppose the changing magnetic flux---corresponding to a steeper potential landscape around the equilibrium position perpendicular to the magnetic field. This brings the amplitude of the oscillations drop to zero. 
We  also note that the  functional form of the  experimental curves depends on the geometry of the crystal, and are not related by a multiplicative factor.

With \(\vec{e} = \vec{e}_1\) and \(\vec{B} = B \vec{e}_3\), to satisfy our  gauge  condition,  $\ve{e} \times \ve{A}(\ve{x}) = 0$, we can choose the Landau gauge $\ve{A}(\ve{x}) = -B(x_2,0,0) $.    In this case,
\begin{equation}
    \frac{\partial\ve{A}(\ve{x})}{\partial x_n} ( \ve{e}\times \ve{x})_n  =  - B \ve{e}_1 (\ve{e}_1 \times \ve{x})_2 =  B \ve{e}_1 x_3
\end{equation}
Since  we  are free  to  choose the origin of the z-axis, we can set $x_3 =0$. Furthermore,  $  \frac{ \partial^2\ve{A}(\ve{x})}{\partial x_n \partial x_m} = 0$.  Therefore, the orbital contribution to $\hat{\tau}$  vanishes, and the imaginary part of the magnetotropic susceptibility is formally given by the same expression as for pure spin systems.  The rationale  for this result is that  for this geometry the temporal  variation of
the magnetic field piercing the 2D  metal is  proportional to  $\Delta\Theta(t)^2$, {\tss thus no eddy-current contribution appears at linear order in $\Delta\Theta(t)$.}
In other words, in linear response, eddy currents  are not  generated by the oscillations of the cantilever.  On the other hand, the total torque $\hat{\tau}$  picks up  contributions up to $\Delta\Theta(t)^2$ 
such that this operator contains an orbital contribution.  To  summarize,  we have for this geometry: 
\begin{eqnarray}
  & &  \hat{\tau}_P =  - g \ve{\hat{S}}_{\text{tot}} \cdot \left(  \ve{e} \times \ve{B} \right)   \\
  & & \hat{\tau}_D = g \ve{\hat{S}}_{\text{tot}} \cdot \ve{e} \times \left( \ve{e} \times \ve{B}  \right)   
+   \frac{B\pi i}{\Phi_0} \sum_{\ve{i},\ve{j}} \hat{\ve{c}}^\dagger_{\ve{i}} h_{\ve{i},\ve{j}}(\ve{A})
     (\ve{j}-\ve{i})_1(\ve{j}+\ve{i})_2 
      \hat{\ve{c}}_{\ve{j}}.  \nonumber
      \label{eq:Metal_B3}
\end{eqnarray}
In the above we have used the short-hand notation $h_{\ve{i},\ve{j}}(\ve{A}) = h_{\ve{i},\ve{j}} e^{\frac{2 \pi i}{\Phi_0} \int_{\ve{i}}^{\ve{j}}  \ve{A}(\ve{x}) \cdot d\ve{x}}$.  
Furthermore, the orbital coupling  of the magnetic field to the electrons in the Hamiltonian $\hat{H}_0$  will  affect the wave functions and spectrum of the system in the form of Landau  levels. 

Fig.~\ref{fig:Metal_PdCoO2} (c),(d) show experimental results for this geometry.  The difference  between the geometry considered in (a) and (b) is striking.  Strictly speaking, damping in a perfect two-dimensional crystal  should only stem from spin fluctuations and should be substantial only in the presence of  low-lying ferromagnetic fluctuations.  In this case we  do not expect  size  effects of  the crystal to kick in after 
normalization by the volume.  Corrections to this ideal picture  seen in Fig.~\ref{fig:Metal_PdCoO2} (d)   may stem from the layered  nature of the crystal  and finite width  as well as non-perfect  alignment  of  the crystal perpendicular to the magnetic field. Both effects can generate eddy current contributions to the torque  operator  $\hat{\tau}_P$. 
It is however apparent that the damping is substantially smaller  than for the geometry considered in Fig.~\ref{fig:Metal_PdCoO2} (a),(b).     The size  dependence of the frequency shift  (see Fig.~\ref{fig:Metal_PdCoO2}(c)) may  arise from the aforementioned deviation from the two-dimensional limit,  as well as from the orbital contribution to $\hat{\tau}_D$ in Eq.~\ref{eq:Metal_B3}.  It is also interesting  to note that in this geometry,  quantum oscillations are observed due  to the 
emergence of Landau  levels in the spectrum of the  system.  We note that the quantum oscillation frequency and the beating are consistent  with the Fermi surface of PdCoO$_2$  \cite{Hicks12}.

Given the above  theoretical modelling, a detailed understanding of  $k(\omega)$ is at reach for  Fermi liquids. In particular, for a  given band structure obtained, for example, from density functional  theory (DFT)~\cite{Hicks12}, an explicit calculation of  $k(\omega)$  is in reach.

\subsection*{Generic one-band model: Rashba spin-orbit coupling versus altermagnetism}

Let us now focus on a metallic system with Rashba spin-orbit coupling (SOC) or altermagnetic term.  We will consider  a  geometry  where the magnetic field is perpendicular to the two-dimensional metal  and assume that it is weak enough to neglect the orbital coupling to the electrons and  associated formation of Landau levels.  Aside from this approximation, the calculation is exact. Specifically, we consider the general Hamiltonian: 
\begin{equation}
  \hat{H} = \sum_{\vec{k}} \hat{c}^\dagger_{\vec{k}} (\epsilon_{\vec{k}} + \vec{v}_{\vec{k}} \cdot \vec{\sigma}) \hat{c}_{\vec{k}} - g \vec{\hat{S}}_{\text{tot}} \cdot \vec{B} = \sum_{\vec{k}} \hat{c}^\dagger_{\vec{k}} (\epsilon_{\vec{k}} + (\vec{v}_{\vec{k}} - g \vec{B}/2) \cdot \vec{\sigma}) \hat{c}_{\vec{k}},
\end{equation}
where \(\hat{c}_{\vec{k}} = (\hat{c}_{\vec{k}, \uparrow}, \hat{c}_{\vec{k}, \downarrow})^T\) and \(\epsilon_{\vec{k}}\) and \(\vec{v}_{\vec{k}}\) are a general dispersion relation and spin-orbit coupling or altermagnetic terms, respectively. If \(\vec{v}_{- \vec{k}} = - \vec{v}_{\vec{k}}\), time-reversal symmetry is conserved, thereby accounting for SOC, whilst if \(\vec{v}_{- \vec{k}} = \vec{v}_{\vec{k}}\), time reversal is broken, thereby accounting for a spin-nematic or altermagnetic Fermi surface.  The \(2 \times 2\) Hamiltonian is given by \(H_{\vec{k}} = \epsilon_{\vec{k}} + \vec{d}_{\vec{k}} \cdot \vec{\sigma}\), with \(\vec{d}_{\vec{k}} = \vec{v}_{\vec{k}} - g \vec{B} / 2\). The eigenvalues are \(\epsilon_{\vec{k}, \pm} = \epsilon_{\vec{k}} \pm \left|\vec{d}_{\vec{k}}\right|\) and the projectors for each eigenstate are \(P_{\vec{k}, \pm} = \frac{1}{2} (1 \pm \frac{\vec{d}_{\vec{k}}}{\left| \vec{d}_{\vec{k}} \right|}  \cdot \vec{\sigma})\).

To calculate the magnetotropic susceptibility, we need to calculate the dynamical spin susceptibility tensor and the total spin. The dynamical spin susceptibility tensor is given by:
\begin{equation}
  \chi^{\alpha\beta}(\omega) = \langle \hat{S}_{\text{tot}}^\alpha \hat{S}_{\text{tot}}^\beta \rangle(\omega) =  \frac{1}{4} \sum_{\vec{k}, \mu, \nu} \text{Tr} \left(\sigma^\alpha P_{\vec{k}, \mu} \sigma^\beta P_{\vec{k}, \nu} \right) \frac{n_F(\epsilon_{\vec{k}, \nu}) - n_F(\epsilon_{\vec{k}, \mu})}{\omega + i0^+ + \epsilon_{\vec{k},\nu} - \epsilon_{\vec{k}, \mu}},
\end{equation}
where \(\mu,\nu = \pm\) and \(n_F(x) = 1 / (e^{\beta x} + 1)\) is the Fermi function. The total spin is given by:
\begin{equation}
  \langle \hat{S}^\alpha_{\text{tot}} \rangle = \frac{1}{2} \sum_{\vec{k}, \mu} \text{Tr} \left(\sigma^\alpha P_{\vec{k}, \mu}\right) n_F(\epsilon_{\vec{k}, \mu}).
\end{equation}
Considering \(g = \text{diag}(g_1, g_2, g_3)\), \(\vec{B} = B \vec{e}_3\) and \(\vec{e} = \vec{e}_2\), the dynamical magnetotropic susceptibility reads
\begin{equation}
  k(\omega) = \frac{1}{N} \left[-  g_3 B \langle \hat{S}_{\text{tot}}^z \rangle + g_1^2 B^2 \chi^{xx}(\omega) \right].
\end{equation}
After some algebra, we arrive at 
\begin{equation}
  k(\omega) = \frac{B}{2N} \sum_{\vec{k}} \frac{h_{\vec{k}}}{\left|\vec{d}_{\vec{k}}\right|} \left[g_3 d_{\vec{k}}^z + \frac{g_1^2 B}{4} \frac{4\left|\vec{d}_{\vec{k}}\right|^2}{(\omega + i0^+)^2 - 4\left|\vec{d}_{\vec{k}}\right|^2} \left(1 - \frac{(d_{\vec{k}}^x)^2 - (d_{\vec{k}}^y)^2 - (d_{\vec{k}}^z)^2}{\left|\vec{d}_{\vec{k}}\right|^2}\right)\right],
\label{k_metal.eq}
\end{equation}
where \(h_{\vec{k}} = \sinh(\beta \left|\vec{d}_{\vec{k}}\right|) / (\cosh(\beta \left|\vec{d}_{\vec{k}}\right|) + \cosh(\beta \epsilon_{\vec{k}}))\) and \(\vec{d}_{\vec{k}} = \vec{v}_{\vec{k}} - \vec{e}_z g_3 B / 2\). 

For \(\vec{v}_{\vec{k}} = 0\) the Fermi surface is metallic and \(\vec{d}_{\vec{k}} = - \vec{e}_z g_3 B / 2\). The dynamical magnetotropic susceptibility is given by:
\begin{equation}
  k(\omega) = \left( \frac{1}{2N} \sum_{\vec{k}} h_{\vec{k}} \right) \left[- g_3 B + \frac{(g_1 B)^2}{2} \left(\frac{1}{\omega + i0^{+} + g_3 B} - \frac{1}{\omega + i0^{+} - g_3 B}\right) \right] 
\end{equation}
At \(\omega = 0\), \(k(\omega = 0) = B \frac{g_1^2 - g_3^2}{g_3} \frac{\sum_{\vec{k}} h_{\vec{k}}(\beta B)}{2N}  \).
So, \(k(\omega = 0) = 0\) for isotropic \(g\) and \(\vec{v}_{\vec{k}} = 0\); furthermore, the scaling \(\beta k(\omega = 0) = f(\beta B)\) is still present as in the free single magnetic moment case. In the same way, the dynamical part has two peaks at \(\omega = \pm g_3 B\), i.e. the Larmor frequency. 

As is apparent from Eq.~\ref{k_metal.eq}, $k''(\omega)$  will show  a continuum of  excitations.  In particular, the frequency $\Omega$  belongs to the support of $k''(\omega)$ if there exists  a $\ve{k}$-vector   in the  Brillouin zone, such that $\Omega = \pm 2 |\ve{d}_{\ve{k}}|$.  This  has  important  consequences.  If  $\ve{v}_{\ve{k}} \cdot \ve{B} =0 $ then  the spectrum  of  excitations is  gapped, the gap  being set by the Zeeman energy: $B g_3$. In this  case no damping of the cantilever oscillations is  expected in the low-frequency regime $\omega < B g_3$.  An example  would  be the  Rashba SOC: $\vec{v}_{\vec{k}} = \lambda (\sin(k_y)\vec{e}_x - \sin(k_x) \vec{e}_y)$. On the  other hand, for a d-wave altermagnetic Fermi surface, $\vec{v}_{\vec{k}} = \Delta(\cos(k_x) - \cos(k_y)) \vec{e}_z$, the spectrum of $k''(\omega)$ can be gapless, if \(\Delta \sim B\), such that damping of the cantilever oscillations is expected in the low-frequency regime. 

On the whole, Eq.~\ref{k_metal.eq} indicates that the simple local-moment scaling relation $\beta k=f(\beta B)$ is generally lost once $\ve{v}_{\vec{k}}\neq 0$.

\subsection*{Magnetotropic response of Kitaev materials}
{\tss Having established the response formalism, we now apply it to spin models relevant to Kitaev materials, with a focus on $\alpha$-RuCl$_3$. The AF-QMC simulations are performed using the Algorithms for Lattice Fermions (ALF)~\cite{ALF_v2.4} implementation, together with an optimized auxiliary-field gauge choice to mitigate the sign problem; details of this technical procedure are given in the Methods section. We note that an  ALF implementation of the spin model in Eq.~\ref{Eq:MMKM}  is  available on the \href{https://alf.physik.uni-wuerzburg.de/Hamiltonians/hamiltonians/kit-heis/readme/}{ALF website}.}
\label{sec:results}

\subsection*{Static magnetotropic scaling of $\alpha$-RuCl$_3$}

The static magnetotropic susceptibility of $\alpha$-RuCl$_3$ was experimentally shown to exhibit a scaling behavior~\cite{Modic20}, $k/T = f(B/T)$, characteristic of a local moment. This scaling behavior was also observed in quantum Monte Carlo simulations of a model for $\alpha$-RuCl$_3$~\cite{SatoT23a}.   Specifically,  the model considered in Ref.~\cite{SatoT23a} 
is defined on the honeycomb lattice and reads:
\begin{eqnarray}
\label{Eq:MMKM}
\hat{H}_{\text{Spin}}&=&\sum_{\langle \ve{i},\ve{j}\rangle_\gamma}\left[K_1\hat{S}_{\ve{i}}^{\gamma} \hat{S}_{\ve{j}}^{\gamma}+\Gamma_1 \left(\hat{S}_{\ve{i}}^{\alpha} \hat{S}_{\ve{j}}^{\beta}+\hat{S}_{\ve{i}}^{\beta} \hat{S}_{\ve{j}}^{\alpha}\right)
+\Gamma'_1 \left(\hat{S}_{\ve{i}}^{\gamma} \hat{S}_{\ve{j}}^{\alpha}+\hat{S}_{\ve{i}}^{\gamma} \hat{S}_{\ve{j}}^{\beta}
+\hat{S}_{\ve{i}}^{\alpha} \hat{S}_{\ve{j}}^{\gamma}+\hat{S}_{\ve{i}}^{\beta} \hat{S}_{\ve{j}}^{\gamma}\right) \right]  \nonumber \\
& & + \sum_{\langle \ve{i},\ve{j}\rangle}  J_{1}  \hat{\ve{S}}_{\ve{i}}  \cdot \hat{\ve{S}}_{\ve{j}} + \sum_{\langle\langle\langle \ve{i},\ve{j}\rangle\rangle\rangle}  J_{3}  \hat{\ve{S}}_{\ve{i}}  \cdot \hat{\ve{S}}_{\ve{j}},
\end{eqnarray}
where $\langle \ve{i},\ve{j} \rangle_\gamma$ refers to nearest-neighbor bonds of type $\gamma=x,y,z$ with cyclic permutations $(\gamma,\alpha,\beta)$ of $(x,y,z)$, $\langle \ve{i},\ve{j} \rangle$ to all nearest-neighbor bonds, and $\langle\langle\langle \ve{i},\ve{j}\rangle\rangle\rangle$ to third-neighbor bonds.
A magnetic field is included through a Zeeman coupling,
\begin{eqnarray}
\label{Eq:HKitaevplusB}
\hat{H}=\hat{H}_{\text{Spin}} +\mu_B\sum_{\ve{i}} g  \hat{\ve{S}}_{\ve{i}} \cdot \ve{B}  ,
\end{eqnarray}
with an anisotropic $g$-tensor consistent with the crystallographic axes of $\alpha$-RuCl$_3$.
In $\alpha$-RuCl$_3$, the orthonormal axes $\mathbf{e}_a$, and $\mathbf{e}_b$, and $\mathbf{e}_c$ are defined by the crystallographic directions, with $\mathbf{e}_c \parallel [111]$, $\mathbf{e}_a \parallel [11\bar{2}]$, and $\mathbf{e}_b \parallel [\bar{1}10]$.
In this convention, $[xyz]$ denote the cubic spin-space axes, $\mathbf{e}_a$ and $\mathbf{e}_b$ lie in the honeycomb plane, and $\mathbf{e}_c$ is perpendicular to it. 
Within the honeycomb plane, $\mathbf{e}_b$ is parallel to one bond direction of the honeycomb lattice, while $\mathbf{e}_a$ is perpendicular to it. 
The magnetic field is applied in the $a$-$c$ plane, nearly parallel to the crystallographic $a$ axis, consistent with the experimental setup.
Unless otherwise stated, we use the parameter set $(J_1, J_3, K_1, \Gamma_1,\Gamma'_1)=(-5.8,5.8,-58,29,0)~{\text K}$ proposed for $\alpha$-RuCl$_3$~\cite{Winter:2017aa}, and $(g_a,g_b,g_c)=(2.3,2.3,1.3)$~\cite{Chaloupka16,Yadav:2016aa}.
In zero field, $\alpha$-RuCl$_3$ exhibits zigzag order below approximately $10\,\mathrm{K}$, while signatures associated with proximity to a Kitaev spin liquid appear at intermediate temperatures up to $\sim 100\,\mathrm{K}$~\cite{Do:2017aa,Banerjee1055}. 

\begin{figure}
\centering
\includegraphics[width=0.9\linewidth]{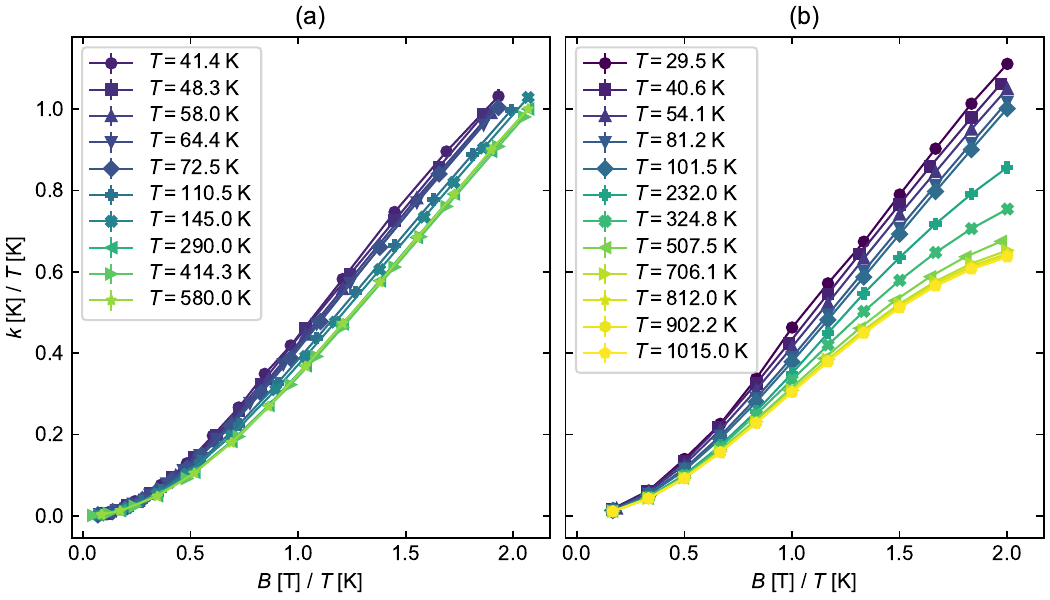}
\caption{\label{fig:Wu_params} 
Static magnetotropic susceptibility of the $\alpha$-RuCl$_3$ model for two parameter sets.
Panel (a) shows the data from Ref.~\cite{SatoT23a} for $(J_1, J_3, K_1, \Gamma_1,\Gamma'_1)=(-5.8,5.8,-58,29,0)~{\text K}$, and panel (b) shows the results for the parameter set proposed in Ref.~\cite{Wu18}, $(J_1, J_3, K_1, \Gamma_1,\Gamma'_1)=(-0.35, 0.34, -2.8, 2.4, 0)\,\mathrm{meV} \approx(-4.1, 3.9, -32.5, 27.9, 0)\,\mathrm{K}$.
 Here $(g_a,g_b,g_c)=(2.3,2.3,1.3)$.
}
\end{figure}

In Ref.~\cite{SatoT23a} we show that at temperatures exceeding the magnetic exchange scale, the susceptibility obeys the scaling form $k=Tf(B/T)$, consistent with independent local moments with anisotropic $g$-factors; the corresponding data are shown in Fig.~\ref{fig:Wu_params} (a).
Upon lowering the temperature below the exchange scale, where the uniform spin susceptibility departs from Curie behavior, a distinct scaling-like collapse is observed.
This behavior is also observed in the pure Kitaev model, whereas it is absent in unfrustrated XXZ models on the honeycomb lattice. In the XXZ model, the breakdown of high-temperature scaling coincides with the buildup of magnetic correlations, and no new collapse is observed at lower temperatures.
The static magnetotropic susceptibility therefore suggests that torque fluctuations in $\alpha$-RuCl$_3$ retain a predominantly local character across a broad temperature window, consistent with a renormalized local-moment picture.

An important question is whether other parameter sets proposed for $\alpha$-RuCl$_3$ also exhibit the same scaling behavior. In Fig.~\ref{fig:Wu_params} (b) we consider the parameter set proposed in Ref.~\cite{Wu18}, \((J_1, J_3, K_1, \Gamma_1, \Gamma_1^\prime) = (-0.35, 0.34, -2.8, 2.4, 0)\,\mathrm{meV} \approx (-4.1, 3.9, -32.5, 27.9, 0)\,\mathrm{K}\), and compute the static magnetotropic susceptibility using unbiased AF-QMC simulations (see Methods). 
Importantly, for this parameter set the ferromagnetic Kitaev interaction \(K_1\) is comparable to the $\Gamma_1$ coupling, and we indeed observe the high-temperature scaling behavior, but not its low-temperature counterpart. {\tss These results are consistent with the renormalized local-moment interpretation proposed previously for Kitaev-dominated regimes. 
At the same time, the absence of the low-temperature collapse for the parameter set proposed in Ref.~\cite{Wu18} shows that the scaling is sensitive to the microscopic parameter set.}

Therefore, a systematic investigation of $k(\omega = 0)$ for parameter sets summarized in, e.g., Ref.~\cite{Moeller25} is an interesting direction for future work that may help further constrain the parameter space of $\alpha$-RuCl$_3$. {\tss More broadly, the robustness of this scaling in other models with short-ranged correlations or different forms of anisotropic exchange remains an important open question.}

\begin{figure}
\centering
\includegraphics[width=0.9\linewidth]{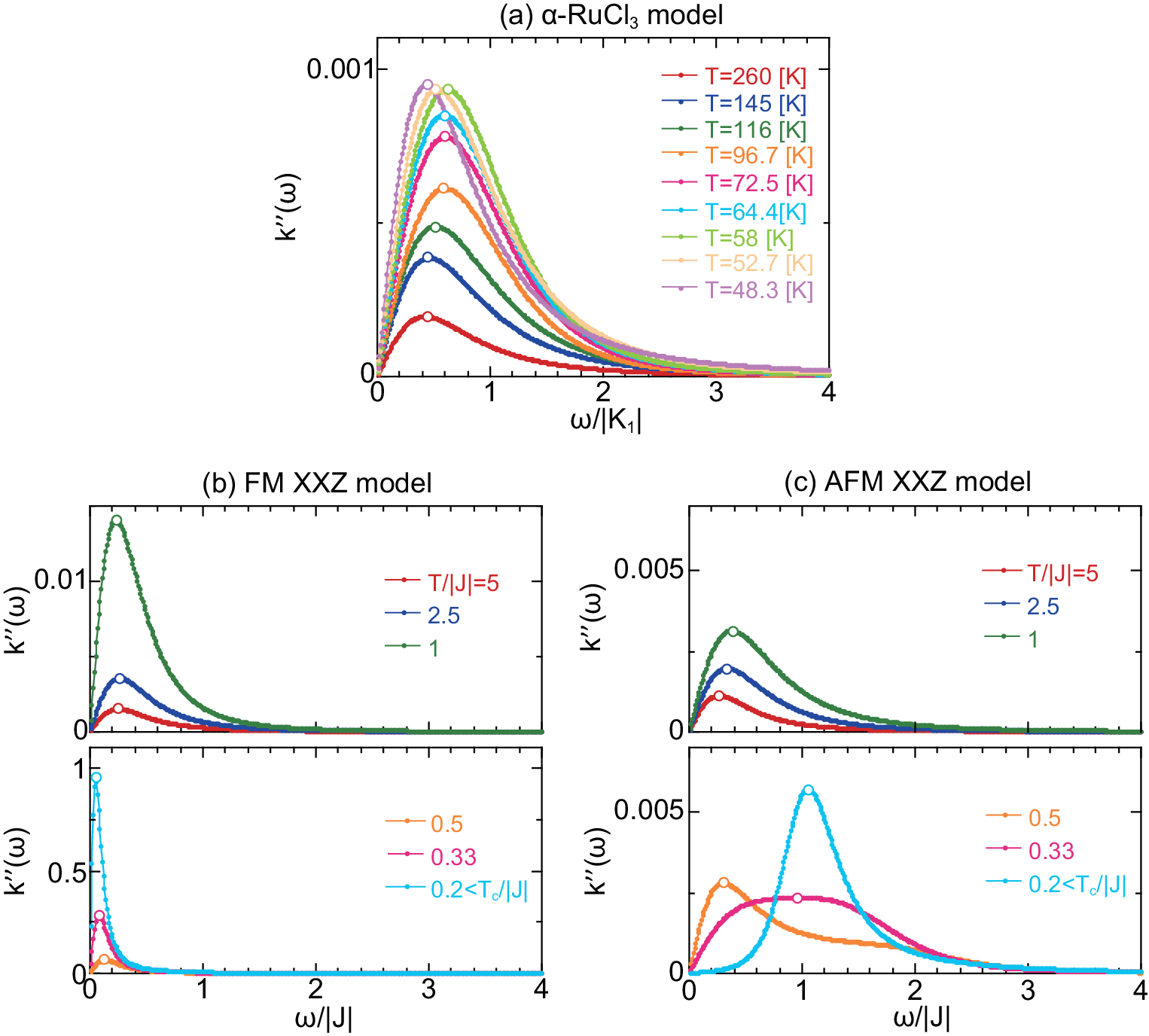}
\caption{\label{fig:k-RuCl3} 
Dynamical magnetotropic susceptibility for the $\alpha$-RuCl$_3$ and XXZ models.
(a) Temperature evolution of $k''(\omega)$ for the $\alpha$-RuCl$_3$ model in a magnetic field $B=5\,\mathrm{T}$ applied in the $a$-$c$ plane, nearly parallel to the crystallographic $a$ axis. 
We use the parameter set $(J_1,J_3,K_1,\Gamma_1,\Gamma'_1)=(-5.8,5.8,-58,29,0)\,\mathrm{K}$ proposed for $\alpha$-RuCl$_3$~\cite{Winter:2017aa}, and $(g_a,g_b,g_c)=(2.3,2.3,1.3)$~\cite{Chaloupka16,Yadav:2016aa}. 
For this parameter set, $k(\omega=0)$ exhibits the $k/T=f(B/T)$ scaling behavior; see Fig.~\ref{fig:Wu_params} (a) and Ref.~\cite{SatoT23a}. 
(b,c) Corresponding results for the non-frustrated XXZ model on the honeycomb lattice with $\mu_B B/|J|=0.1$: (b) ferromagnetic case, $(J,J_z)=(-1.0,0.5)$, and (c) antiferromagnetic case, $(J,J_z)=(1.0,-0.5)$. 
The magnetic field is applied along the  $\mathbf{e}_c$ ([001] direction in the cubic spin basis), with $\mathbf{e}\parallel \mathbf{e}_b$ ([010]). 
The $g$ tensor is taken diagonal with $(g_a,g_b,g_c)=(2,2,1)$. 
Open circles indicate the positions of the spectral maxima.
}
\end{figure}

\subsection*{Dynamical magnetotropic susceptibility of the model for $\alpha$-RuCl$_3$}

The above results for the static magnetotropic susceptibility suggest that the torque fluctuations in $\alpha$-RuCl$_3$ retain a predominantly local character across a broad temperature window, consistent with a renormalized local-moment picture. As shown in the theoretical framework introduced above,
the imaginary part of $k(\omega)$ relates to the uniform, $\ve{q}=0$,  transverse dynamical spin  susceptibility that is expected to pick up fractionalization in terms of a broad continuum of excitations. Using the  ALF \cite{ALF_v2.4} implementation of the stochastic analytical continuation method \cite{Beach04a,Sandvik98,Shao23} we can compute  $k''(\omega)$ from the  imaginary time AF-QMC data to obtain a complete understanding of the complex function $k(\omega)$.

Figure~\ref{fig:k-RuCl3} (a) shows the temperature evolution of the dynamical magnetotropic spectrum $k''(\omega)$ for the $\alpha$-RuCl$_3$ model in a magnetic field of $5\,\mathrm{T}$, within the temperature range accessible to our simulations.
At the highest temperatures,  $k''(\omega)$ exhibits a simple single-peak structure at a frequency of order the Zeeman scale $\mu_B g_{\mathrm{eff}} B$, consistent with a local-moment-like response (see Eq.~\ref{eq:local_moment_k}). Upon lowering the temperature, the overall spectral intensity increases, while the line shape remains largely unchanged and retains a single-peak structure.
The peak position displays only a weak and slightly non-monotonic temperature dependence, with no substantial overall shift. {\tss While $k''(\omega)$ for a local moment shows a $\delta$-function peak at the Larmor frequency,  our numerical data show a broad asymmetric single peak feature. Heuristically,  we can account for  this broadening by  replacing  $i 0^{+} $ in  Eq.~\ref{eq:local_moment_k} with a self-energy term,  $i \Sigma''(\omega)$.   Causality requires that $\Sigma''(\omega)$ be an odd function of $\omega$, such that this heuristic form remains   consistent with the scaling behavior of the magnetotropic susceptibility.  We have seen that $k''(\omega)$ is related to the $\vec{q}=0$ dynamical spin susceptibility  (see Eq.~\ref{eq:imag_k}). For  Kitaev models, in the  zero temperature   limit and in the absence of a magnetic field, this quantity is known to show a broad asymmetric continuum of excitations above the flux  gap \cite{Knolle14}  the  width being set by the Kitaev exchange scale. 
We also note that the spectral function of Ref.~\cite{Knolle14} shows little momentum dependence, a picture that is consistent with the local-moment interpretation. Clearly, the frequency scale considered in Fig.~\ref{fig:k-RuCl3} (a) is orders of magnitude larger than   experimentally achievable in RTM experiments. Nevertheless, it shows that  since at the experimentally relevant frequency scales, $k''(\omega)$  is  very  small and no prominent/anomalous damping is expected. This is consistent with experimental observations \cite{Modic20}.  From the theoretical point of view the  scaling of  the magnetotropic susceptibility  as well as the functional form of $k''(\omega)$    place some constraints  on  $k(\omega)$  and are hence of importance for future theoretical considerations.}

\subsection*{Comparison to an unfrustrated XXZ model}
As a benchmark, we compare these results to those obtained for an unfrustrated XXZ model on the honeycomb lattice, described by 
$\hat{H}_{\text{Spin}}=\sum_{\langle \ve{i}, \ve{j} \rangle}  J ( \hat{S}^x_{\ve{i}}  \hat{S}^x_{\ve{j}} +  \hat{S}^y_{\ve{i}}  \hat{S}^y_{\ve{j}} )
+ \left(J + J_{z} \right) \hat{S}_{\ve{i}}^{z} \hat{S}_{\ve{j}}^{z} $.
We present quantum Monte Carlo results for $J_z/J = -0.5$
 and $\mu_B B/|J| = 0.1$. 
Details of the finite-temperature phase diagram for the parameter sets $(J,J_z)=(-1.0,0.5)$ and $(1.0,-0.5)$ are provided in the Supplemental Information.

Figure~\ref{fig:k-RuCl3} shows $k''(\omega)$ for the unfrustrated XXZ model. 
At high temperatures, the spectrum exhibits a single peak, similar in overall appearance to the \(\alpha\)-RuCl$_3$ case.
Upon lowering the temperature, however, the spectral evolution differs qualitatively from that of the \(\alpha\)-RuCl$_3$ model.
In the ferromagnetic (FM) XXZ case (see Fig.~\ref{fig:k-RuCl3} (b)), lowering the temperature drives a pronounced transfer of spectral weight towards $\omega \to 0$. 
The peak progressively shifts to lower frequencies and its amplitude increases strongly, consistent with the build-up of low-energy magnetic fluctuations near the magnetic transition.
In the antiferromagnetic (AFM) XXZ case (see Fig.~\ref{fig:k-RuCl3} (c)), the response does not soften toward $\omega\to 0$.
Instead, the dominant spectral weight remains at finite frequencies of order $J$, and the very-low-frequency response is comparatively weaker than in the ferromagnetic case.
This indicates that the dominant magnetic response remains at finite energies rather than shifting toward $\omega \to 0$.
In contrast to these two distinct XXZ trends---softening toward $\omega \to 0$ in the FM case and a finite-frequency response in the AFM case---the \(\alpha\)-RuCl$_3$ spectra exhibit only a weak temperature dependence of their overall structure (Fig.~\ref{fig:k-RuCl3} (a)).
Upon lowering the temperature, the overall spectral intensity increases, while the spectrum retains a broad single-peak structure and the peak position changes only weakly.
Thus, the \(\alpha\)-RuCl$_3$ response does not show a clear tendency toward either of the two XXZ cases.

To elucidate these differences, we recall that $k(\omega)$ probes spin fluctuations transverse to the applied magnetic field. 
The contrasting spectral evolutions therefore reflect qualitatively different transverse spin dynamics in the XXZ and \(\alpha\)-RuCl$_3$ models. 
For the XXZ geometry used in Fig.~\ref{fig:k-RuCl3}, we take $\mathbf{B}\parallel \mathbf{e}_c$ and $\mathbf{e}\parallel \mathbf{e}_b$ in the cubic spin basis.  
With a diagonal $g$ tensor, one obtains $g^T(\mathbf{e}\times\mathbf{B})\parallel \mathbf{e}_a$, such that $k''(\omega)$ projects the transverse response $\chi^{xx}{}''(\omega)$ up to an overall prefactor [Eq.~\ref{eq:imag_k}]. 
The corresponding $\chi^{xx}{}''(\omega)$, shown in the Supplemental Information, exhibits the same temperature evolution as $k''(\omega)$, confirming that the spectral redistribution in Fig.~\ref{fig:k-RuCl3} reflects the transverse spin dynamics probed in this geometry.

In contrast to the XXZ geometry, the field direction used for the \(\alpha\)-RuCl$_3$ model does not align with a principal axis of the cubic spin basis.
Consequently, the vector $g^T(\mathbf{e}\times\mathbf{B})$ contains several components, such that $k(\omega)$ probes a linear combination of spin susceptibilities $\chi^{\alpha\beta}(\omega)$.
Moreover, the bond-dependent Kitaev and off-diagonal $\Gamma$ interactions give rise to anisotropic and off-diagonal spin correlations.
The dynamical magnetotropic response therefore cannot be reduced to a single spin component, but instead reflects a mixture of transverse spin fluctuations.
Accordingly, within the temperature range accessible to our simulations, the \(\alpha\)-RuCl$_3$ spectra do not show a clear tendency toward either of these two XXZ behaviors.
Rather, the overall spectral intensity increases upon lowering the temperature, while the spectrum retains a broad single-peak structure and the peak position changes only weakly.

\begin{figure}
\centering
\includegraphics[width=0.9\linewidth]{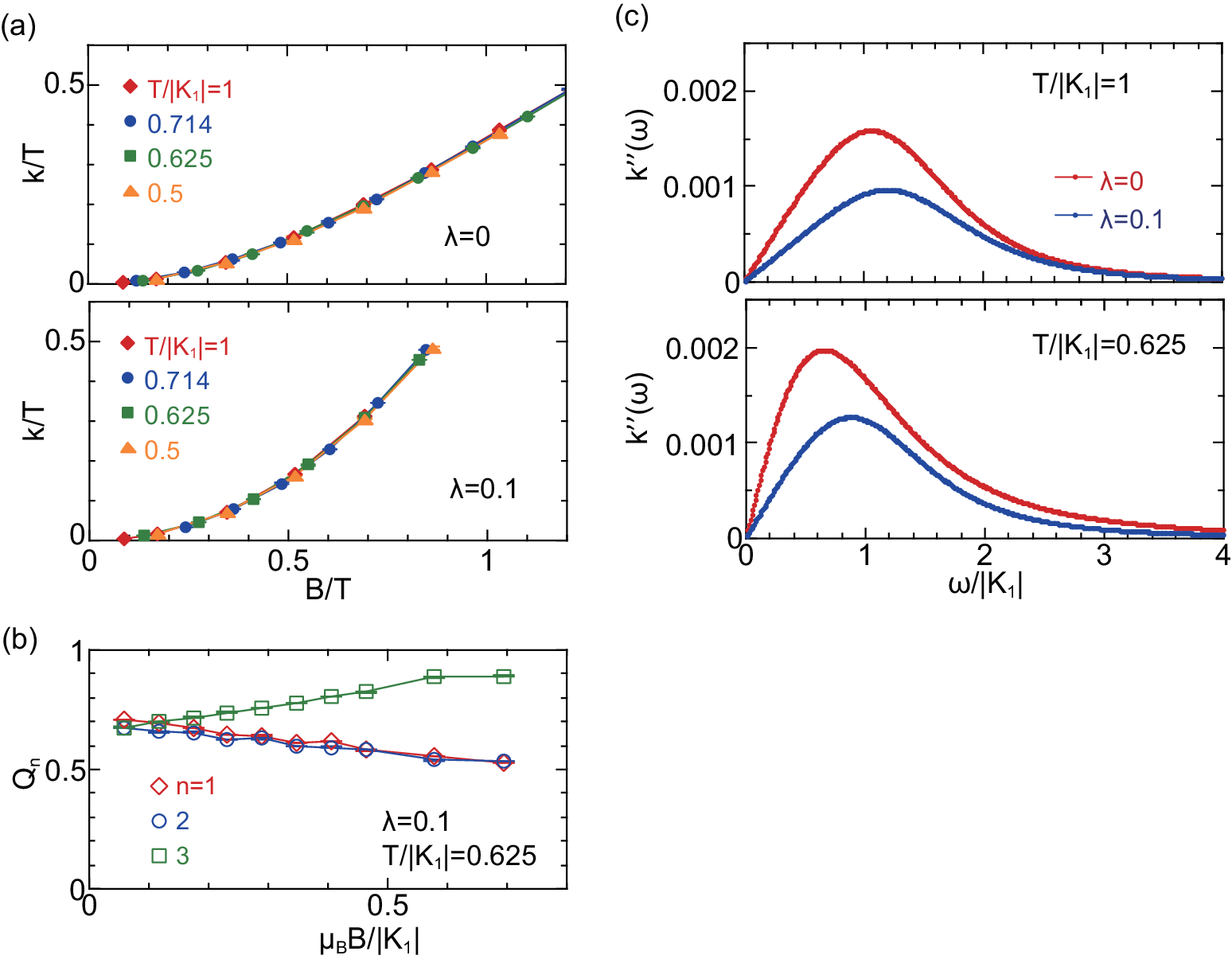}
\caption{\label{fig:k-kitaevphonon} 
{\tss Optical-phonon effects on the Kitaev-limit magnetotropic response.}
(a) Static magnetotropic susceptibility normalized by temperature, $k/T$, as a function of magnetic field normalized by temperature, $B/T$, for the Kitaev limit ($\Gamma_1 = \Gamma'_1=J_1 = J_3 = 0$) of our $\alpha$-RuCl$_3$ model. 
The upper panel shows the pure Kitaev model without spin-phonon coupling ($\lambda=0$), while the lower panel includes coupling to optical phonons with $\lambda=0.1$ and $\omega_0=0.75$. 
In both cases, data at different temperatures collapse onto a single curve as a function of $B/T$, demonstrating the robustness of the $k/T$-$B/T$ scaling below the magnetic energy scale.
(b) Magnetic-field dependence of the bond phonon displacement $Q_b$ for $\lambda=0.1$. 
$Q_1$, $Q_2$, and $Q_3$ couple to the Kitaev terms $S_x S_x$, $S_y S_y$, and $S_z S_z$, respectively, 
and we find $Q_1 = Q_2 \neq Q_3$.
(c) Dynamical magnetotropic susceptibility $k''(\omega)$ in the Kitaev limit, comparing the pure Kitaev case ($\lambda=0$) and the spin-phonon coupled case ($\lambda=0.1$, $\omega_0=0.75$) at two representative temperatures. {\tss Here \(\mu_B B / |K_1| = 0.0579\).}
The coupling to optical phonons suppresses the overall spectral intensity across the frequency range shown. 
}
\end{figure}

\subsection*{Effect of optical phonons}
In real materials, lattice degrees of freedom are inevitably present and can influence magnetic response functions \cite{sandilands_scattering_2015,li_giant_2021}. It has been shown by theory and experiments that phonon coupling in Kitaev spin liquids is strong at fractional excitation energy scales \cite{feng_footprints_2022,li_giant_2021,Shragai2026aa}. 
That is, near the flux gap and \(K\) energy scales, phonons become soft and destabilise the spin liquid. The question now is whether spin-phonon interactions can renormalise the $k/T$-$B/T$ scaling and dynamical features of the magnetotropic susceptibility, both associated with a local moment picture due to strong Kitaev interactions. Then, 
to assess the role of optical phonons in the magnetotropic susceptibility, we consider a minimal spin-phonon model in which $S=1/2$ spins are coupled to optical phonon modes.
We therefore first consider the Kitaev limit of our $\alpha$-RuCl$_3$ model with $\Gamma_1 =\Gamma'_1= J_1 = J_3 = 0$, which serves as a clean reference point for assessing phonon-induced modifications of the magnetotropic response.

Within the Kitaev-limit setting introduced above, we incorporate optical phonons by allowing the bond-dependent Kitaev exchange to couple linearly to lattice displacements. 
The resulting spin-phonon Hamiltonian takes the form
$\hat{H}_{\text{Spin}} = \sum_{b=\langle \ve{i},\ve{j}\rangle_\gamma} \left(1 + g \hat{Q}_b\right) K_1 \hat{S}_{\ve{i}}^\gamma \hat{S}_{\ve{j}}^\gamma + \sum_{b=\langle \ve{i},\ve{j}\rangle} \left( \frac{\hat{P}_b^2}{2m} + \frac{k}{2} \hat{Q}_b^2 \right)$,
where $b=\langle \ve{i},\ve{j}\rangle_\gamma$ denotes nearest-neighbour bonds of type $\gamma = x,y,z$. 
The phonon displacement $\hat{Q}_b$ and conjugate momentum $\hat{P}_b$ reside on each bond and satisfy $[\hat{Q}_b,\hat{P}_{b'}] = i\delta_{bb'}$.
We solve this model using the unbiased AF-QMC method established in our previous work (Ref.~\cite{Inacio25}, see Methods), enabling a numerically exact treatment of the coupled spin-phonon degrees of freedom.
Unless otherwise specified, we consider spin-phonon coupling $\lambda \equiv \frac{g^2}{2k} = 0.1$ and  frequency $\omega_0  \equiv \sqrt{\frac{k}{m}}= 0.75$ in the following. 

To investigate the impact of the spin-phonon coupling on the magnetotropic response, we first examine the $k/T$-$B/T$ scaling.
Figure~\ref{fig:k-kitaevphonon} (a) compares the pure Kitaev case ($\lambda=0$) with the spin-phonon model ($\lambda=0.1$, $\omega_0=0.75$). 
As established in our previous work~\cite{SatoT23a}, the pure Kitaev model exhibits a robust $k/T$-$B/T$ scaling below the magnetic energy scale set by $|K_1|$. {\tss This scaling collapse persists for the spin-phonon coupling considered here.
The data for $\lambda = 0.1$ continue to collapse onto a single curve as a function of $B/T$, showing that the scaling structure of the magnetotropic response is not destroyed by this coupling.}
At the same time, the scaling function itself is modified. 
While the collapse remains intact, the functional form $f(B/T)$ is renormalized in the presence of phonons, reflecting the feedback of bond phonon fluctuations on the magnetic response.

The magnetic-field dependence of the bond phonon displacement shown in Fig.~\ref{fig:k-kitaevphonon} (b) reveals that the lattice degrees of freedom actively respond to the anisotropic spin correlations induced by the field. 
The relation $Q_1 = Q_2 \neq Q_3$ is consistent with the symmetry of the applied field and the bond-selective structure of the Kitaev interactions. 
This suggests that part of the magnetic anisotropy is transferred to the lattice sector, providing a natural microscopic origin for the renormalization of the scaling function observed in Fig.~\ref{fig:k-kitaevphonon} (a).

Figure~\ref{fig:k-kitaevphonon} (c) compares the dynamical magnetotropic spectrum $k''(\omega)$ in the Kitaev limit with and without spin-phonon coupling at \(\mu_B B /|K_1| = 0.0579\). 
For the parameters shown, coupling to optical phonons suppresses the overall spectral intensity across the measured frequency window. 
This suppression indicates that the anisotropy-induced torque fluctuations are reduced overall.
One possible interpretation is that the coupling to optical phonons renormalizes the effective spin interactions through bond phonon fluctuations, thereby reducing anisotropy-driven torque fluctuations and leading to the observed suppression of the dynamical spectral weight probed by $k''(\omega)$.

{\tss
\subsection*{Interpretation of the scaling behaviour}
One key aspect of the Kitaev model is that spin-spin correlations are restricted to nearest-neighbor bonds  due to  flux conservation \cite{baskaran_exact_2007}  such that the magnetic correlation length  is bounded by the lattice spacing. 
This is in stark contrast to a generic spin model, where the magnetic correlation length can diverge at low temperatures for  critical and ordered phases,  and is finite for  spin-gapped states.   Our  numerical calculations 
on a selected set of models suggest that  the observed low-temperature scaling of the magnetotropic susceptibility is a characteristic feature of the Kitaev model. This scaling is equally observed in the simple case of a local moment,  where the magnetic correlation length vanishes.  In contrast, for a  mean-field parton  description of  a  U(1)  Dirac spin liquid, where  the magnetic correlation length diverges, low-temperature scaling is not observed, see Fig.~\ref{fig:spin_liquid}. We also do not observe  low-temperature scaling  for the XXZ model  as well as for the parameter set proposed in Ref.~\cite{Wu18} for $\alpha$-RuCl$_3$, where the Kitaev coupling is not dominant.  
Our  calculations further show that if  the aforementioned flux-conservation symmetry is \textit{weakly} broken,  by including \textit{small} perturbations in comparison to the Kitaev exchange interactions and  magnetic fields, the scaling of the magnetotropic susceptibility is preserved.

From  these   observations we  conjecture that the low-temperature scaling of the magnetotropic  susceptibility is not a signature of  fractionalization, but rather a consequence of a magnetic correlation length  set by  the lattice spacing. Since this is a characteristic feature  of the Kitaev model, we  further conjecture  that the observation of scaling  is a property of models with dominant Kitaev interactions. Future work will  be required to prove or disprove this conjecture.

An understanding of $k(\omega)$ in terms of a local moment model is not tenable for the Kitaev model. In fact, since  $k''(\omega)$ reflects the transverse $\vec{q} = 0$ dynamical spin susceptibility, it is expected to pick up fractionalization in terms of a broad continuum of excitations. Hence  $k(\omega)$ for the  Kitaev model  picks up both fractionalization  and the local nature  of the real-space spin correlations.}

\section*{Discussion}
Our analysis reveals that the dynamical magnetotropic susceptibility probes ultra-low-energy, $\vec{q} = 0$ charge and spin fluctuations. As such it provides a  new  tool  to study quantum critical points (QCP) involving  $\ve{q}=0$ fluctuations. In particular, at these points, fluctuations are scale free and  extend  to $T=0$  and $\omega=0$.   In  what follows we will illustrate this  with 
two examples:  Kondo destruction (KD) quantum criticality in heavy fermion compounds \cite{Schroeder00,Si01,Mazza24}, and the field-induced metamagnetic transition in UTe$_2$ \cite{Zambra25}.  In the latter case,  we provide experimental results.  

The coupling to the charge degrees of freedom stems from the formation of eddy currents that damp the oscillations of the cantilever. Eddy currents  have  many similarities to the  optical 
conductivity $\sigma(\omega)$.  They are  generated only  if  the conductivity is finite.  However, the conductivity is a bulk material property  while eddy currents  are a
macroscopic phenomenon that depends on the size of the sample.  

Mott  quantum criticality  \cite{Imada_rev}  involves the localization of charge degrees of freedom: the metallic phase has low-lying spin and charge fluctuations, while the insulating phase has only {\tss low-lying} spin excitations.  Across this transition, we hence  expect to observe a very strong change in the damping of the cantilever oscillations. 

We conjecture  that a  complete gapping out of the Fermi surface is not necessary to observe a strong change in the damping of the cantilever. An orbital-selective Mott transition \cite{Vojta10} in which
one subset of electrons localizes, thereby changing the Fermi surface volume, should also lead to a strong change in the damping of the cantilever oscillations.
Kondo destruction (KD) quantum criticality \cite{Paschen21} is nothing but a realization of an orbital-selective Mott transition \cite{Senthil03,Danu20,Pan25}.  It separates two distinct ground states: one with static Kondo screening and a large Fermi surface comprising both itinerant and localized electrons (which carry local moments), and one without Kondo screening and a small conduction-electron-only Fermi surface. Critical fluctuations in both the local-moment and charge sectors should emerge from such a KD QCP. As they arise from KD, they are (essentially) local in real space and thus extended in momentum space, i.e., with response also at $\vec{q} = 0$. Furthermore, KD quantum criticality is scale-free, and thus the fluctuations extend to $T=0$ and $\omega=0$. 

\begin{figure}
\centering
\includegraphics[width=\linewidth]{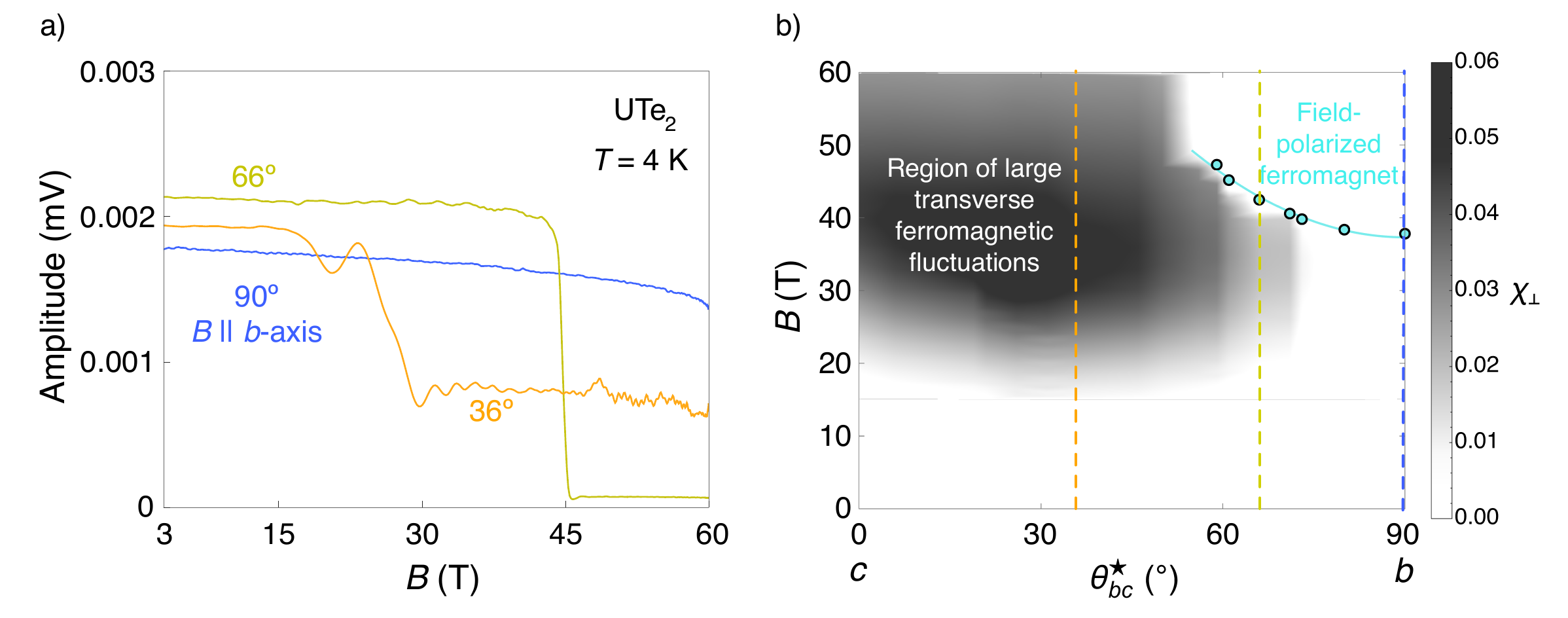}
\caption{Transverse ferromagnetic fluctuations in UTe$_2$. (a) Amplitude of the pickup voltage of the cantilever during magnetotropic susceptibility measurements of UTe$_2$ at 4 K, with the magnetic field applied at different angles in the $bc$ plane, taking 0$^\circ$ for $B \parallel c$. (b) Phase diagram of the transverse susceptibility of UTe$_2$ in the $bc$ plane, obtained using the same technique, data are taken from Ref.~\cite{Zambra25}. The gray region indicates where transverse ferromagnetic fluctuations dominate, while the cyan line and dots correspond to the metamagnetic transition. Colored dashed lines indicate the sample angles shown in panel (a), with their corresponding colors. As the field is oriented toward the $c$ axis, the amplitude drop is consistent with the region of strong transverse ferromagnetic fluctuations.}
\label{fig:Ferromagnetic_fluctuaitons} 
\end{figure}

While the  above discussion highlights the  coupling to charge degrees of freedom, a concrete realization of these ideas in the spin sector is provided by recent magnetotropic measurements in UTe$_2$, where transverse, ultra-low-energy fluctuations emerge in the vicinity of a field-induced transition. As shown in Fig.~\ref{fig:Ferromagnetic_fluctuaitons} (a), the amplitude of the cantilever oscillations exhibits a pronounced suppression at specific field orientations, while Fig.~\ref{fig:Ferromagnetic_fluctuaitons} (b) maps out the corresponding phase diagram of the transverse susceptibility in the bc plane. For magnetic field applied along the b-axis, the transition into the field-polarized state is strongly first order 
\cite{Aoki19,Ran19,Tokiwa24,Tokunaga23}, and the amplitude remains essentially unchanged despite the jump in the frequency shift  \cite{Zambra25}. In contrast, as the field is rotated toward the c-axis, this transition becomes progressively more continuous, approaching a second-order critical end point where fluctuations are strongly enhanced \cite{Wu25}.
The dramatic drop observed in the oscillation amplitude coincides with the region of large transverse ferromagnetic fluctuations, previously identified as a large decrease in the resonant frequency. This behavior reflects the onset of strong dissipation arising from low-energy fluctuations in the vicinity of the critical end point, where the metamagnetic transition terminates. The suppression of the amplitude therefore provides direct evidence that transverse ferromagnetic fluctuations dominate the low-energy dynamics and couple efficiently to the cantilever motion, highlighting the unique capability of dynamical magnetotropic susceptibility to access critical fluctuations at $\ve{q}=0$ in both spin and charge sectors.

\label{Sec:outlook}

{\tss In summary, we have provided a microscopic framework for understanding the dynamical magnetotropic susceptibility $k(\omega)$ in interacting lattice models. 
Our analysis builds on the earlier work of Ref.~\cite{Shekhter23} and extends it by accounting for charge degrees of freedom alongside spin fluctuations. 
This formulation provides a common linear-response framework for localized spin systems and metallic states and is directly suitable for numerical many-body calculations.}

For quantum spin systems, the static magnetotropic susceptibility, $k(\omega=0)$, responds solely to magnetic anisotropy, while the imaginary part captures spin fluctuations transverse to the magnetic field regardless of the magnetic anisotropy. Hence, damping of the cantilever oscillations signals low-lying ferromagnetic fluctuations, indicating proximity to a ferromagnetic instability as  observed in UTe$_2$ {\tss and discussed above};
see Ref.~\cite{Zambra25}. 

Quantum Monte Carlo methods excel at computing susceptibilities, since no analytical continuation is required. Hence, an accurate 
estimate for the $k(\omega = 0 )$  can be obtained from  bias free model  calculations.  For frustrated spin systems, however, these approaches are hampered by the negative sign problem, which limits access to low temperatures and large system sizes. 
However, spin exchange constants are typically on the meV scale, such that by optimizing the severity of the sign problem one can reach relevant temperature windows.
More precisely, the spin model used to account for experimental results is defined on the meV scale set by the dominant exchange interaction $J$; stabilizing simulations at, say, $\beta J \simeq 3$ (where $\beta$ is the inverse temperature) already suffices to access  relevant temperature scales.
{\tss In the simulations presented here, we mitigate the sign problem by using an optimized auxiliary-field gauge choice, whose technical details are described in the Methods section. 
The same AF-QMC formulation also allows us to include couplings to optical phonons without substantially worsening the sign problem.}

{\tss Combining measurements on $\alpha$-RuCl$_3$ with model calculations, we find that the low-temperature scaling behavior of $k(\omega = 0)$ is reproduced in Kitaev-dominated parameter regimes. 
This view is supported by numerical calculations for the parameter set of Ref.~\cite{Wu18}, in which the Kitaev and $\Gamma$ couplings are comparable and the low-temperature scaling is not observed, whereas the pure Kitaev model does exhibit scaling. 
These results are consistent with a renormalized local-moment picture and suggest that the scaling can serve as one diagnostic constraint on candidate microscopic models for $\alpha$-RuCl$_3$.}

Our simulations also give access to $k''(\omega)$, and we have confirmed that for the model parameters reproducing the low-temperature scaling of the magnetotropic susceptibility, $k''(\omega)$ shows a broad feature consistent with the notion of fractionalization.
We have furthermore confirmed that coupling to optical phonons does not destroy the low-temperature scaling behavior of $k(\omega=0)$.

The article provides the theoretical framework for the magnetotropic susceptibility of metallic systems, where both charge and spin degrees of freedom govern the response. The charge contribution is dominated by eddy currents, which damp the time variation of the magnetic flux through the metal. For quasi-two-dimensional materials, this effect depends on the orientation of the magnetic field relative to the metal and can therefore be tuned. This was explicitly shown by measurements on PdCoO$_2$.
Of particular interest will be to investigate Mott or orbital-selective Mott transitions with this method. The localization of a subset of charge carriers can potentially produce a strong change in the magnetotropic susceptibility, since localization suppresses the orbital coupling to the magnetic field and enhances the relative importance of the Zeeman energy. Continuous orbital-selective Mott transitions are equivalent to Kondo-destruction transitions, realized in materials such as YbRh$_2$Si$_2$~\cite{Paschen04} and in model calculations~\cite{Danu20}. It will be interesting to investigate whether the magnetotropic susceptibility can be used to 
study such transitions.

 Measurements of the  magnetotropic susceptibility are versatile; they can be performed on nanogram-sized crystals, they have been demonstrated in magnetic fields up to 65 T, and can be carried out at dilution refrigerator temperatures. While current measurements are performed in the kHz frequency range, this range is only limited by the oscillation frequency of the cantilever. Future developments can push this up to the few GHz range by miniaturisation and use of focused ion beam lithography to extract small, pristine samples. This will allow access to more interesting dynamics, particularly for gapless excitations in quantum spin liquids.
 
 To conclude, the magnetotropic susceptibility provides a powerful probe of the low-energy uniform properties of quantum materials, and the interplay between model calculations and measurements holds the potential to shed light on a broad range of phenomena.

\backmatter

\section*{Methods}

{\tss
\subsection*{PdCoO$_2$ crystal synthesis}

The PdCoO$_2$ parent crystal was provided by Andrew P.~Mackenzie and
Seunghyun Khim. It was grown using the metathetical-reaction method
described by Tanaka \textit{et al.}~\cite{Tanaka1996}. The orientation of the crystals was determined via their growth edges, which are oriented perpendicular to the crystallographic axes.

\subsection*{PdCoO$_2$ sample preparation and characterization}

The larger sample was measured in its original form and was not mechanically cut or polished before the measurements. Its crystallographic orientation was determined using electron backscatter diffraction (EBSD), as shown in Fig.~\ref{fig:ebsd}. The EBSD analysis indexed $100\%$ of the measured points as PdCoO$_2$,
with a mean angular deviation of $0.18^{\circ}$. The indexed $a$- and $b$-axis directions were marked and subsequently used to control the sample orientation.

\begin{figure}[tbp]
    \centering
    \includegraphics[width=1\textwidth]{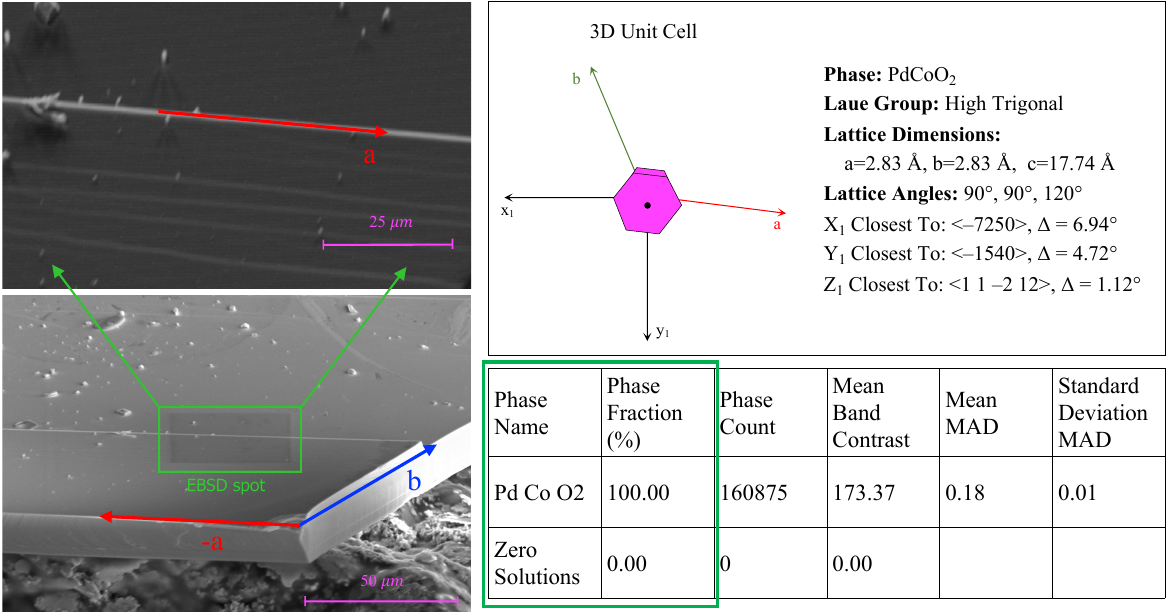}
   \caption{
{\tss EBSD characterization of the parent PdCoO$_2$ crystal. 
The electron-microscopy images on the left show the crystal surface and the region selected for EBSD analysis. 
The indexed crystallographic $a$ and $b$ directions are indicated by red and blue arrows, respectively. 
The upper-right panel shows the corresponding three-dimensional unit-cell orientation with respect to the microscope coordinate system, while the lower-right panel confirms that the analyzed region is identified as PdCoO$_2$ with an approximately $100\%$ phase fraction.}}
    \label{fig:ebsd}
\end{figure}

A smaller sample was prepared from the large crystal using a
Helios G4 xenon plasma focused-ion-beam (FIB) system, as shown in Fig.~\ref{fig:fib}. The sample was mounted with the c-axis perpendicular to the surface of the sample holder and cut using small currents ($4-15 ~\mathrm{nA}$) and $25~\mathrm{kV}$ to minimize surface damage.

The large and small samples, therefore, are taken from the same bulk crystal with the same crystallographic alignment, allowing a direct comparison between their magnetotropic and quantum-oscillation responses.

\begin{figure}[tbp]
    \centering
    \includegraphics[width=0.55\textwidth]{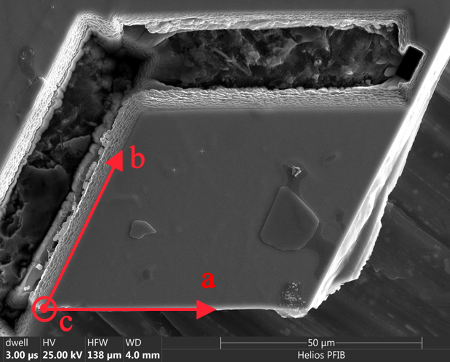}
    \caption{{\tss Scanning-electron-microscopy (SEM) image showing the preparation of the small PdCoO$_2$ sample using FIB milling.
The crystallographic orientation of the FIB'ed sample was defined using the EBSD characterization of the parent crystal.}}
\label{fig:fib}
\end{figure}

The dimensions and estimated masses of the samples are summarized in
Table~\ref{tab:pdco2_samples}. The masses were estimated from the microscopy-derived volumes
using a PdCoO$_2$ density of
$\rho \simeq 7.99~\mathrm{g\,cm^{-3}}$.

\begin{table}[tbp]
\centering
\caption{
{\tss Summary of the PdCoO$_2$ samples used in this work.}}
\label{tab:pdco2_samples}
\begin{tabular}{lcccc}
\hline\hline
Sample
& Origin
& Surface area [$\mu\mathrm{m}^{2}$]
& Volume [$\mu\mathrm{m}^{3}$]
& Estimated mass
\\
\hline
Large
& Parent crystal
& $5.5\times10^{4}$
& $7.5\times10^{5}$
& $5.99~\mu\mathrm{g}$
\\
Small
& FIB-cut from large sample
& $3575$
& $5.0\times10^{4}$
& $0.400~\mu\mathrm{g}$
\\
\hline\hline
\end{tabular}
\end{table}

\subsection*{PdCoO$_2$ magnetotropic measurements}

Magnetotropic susceptibility measurements were performed using resonant torsion magnetometry in the Modic Laboratory at the Institute of Science and Technology Austria \cite{Modic2018}. All measurements were carried out in a $14~\mathrm{T}$ Oxford Instruments Teslatron superconducting magnet over the temperature range $1.7$--$80~\mathrm{K}$. The magnetic field was swept between $0$ and $14~\mathrm{T}$ at a rate of $0.2~\mathrm{T\,min^{-1}}$. Combining the EBSD, mounting, probe-alignment,
and rotator uncertainties gives an estimated absolute angular uncertainty
of approximately $3^{\circ}$--$5^{\circ}$.

The measurements were performed using commercially available atomic force microscopy (AFM) tips called an Akiyama probe \cite{Akiyama2010}. The probe consists of a silicon microcantilever coupled to a quartz tuning fork, which provides electrical excitation and detection of the mechanical resonance. The tuning-fork--cantilever assembly was mounted on a G10 substrate attached to the stage of an Attocube rotator, which was used to control the equilibrium position of the cantilever relative to the external magnetic field. The probe was inserted into a vacuum-walled stainless-steel insert, and the probe space was evacuated to approximately $p \simeq 10^{-5}~\mathrm{mbar}$. Subsequently, approximately $1~\mathrm{mbar}$ of $^4$He exchange gas was introduced. The insert was then placed in a liquid-helium cryostat.

The samples were attached to the end of the cantilever using a small amount of Apiezon L grease. The crystallographic $c$-axis was aligned approximately normal to the cantilever surface, while the crystallographic $b$-axis was aligned approximately along the length of the lever. Optical images of the mounted samples are shown in Fig.~\ref{fig:pdco2_samples}. The sample positions were inspected after the measurements, confirming that no displacement had occurred during the field sweeps and thermal cycling.

The large sample was measured in two field-orientation configurations,
with the magnetic field oriented approximately parallel and perpendicular
to the cantilever. Measurements were performed at $1.7$, $10$,
$40$, $60$, and $80~\mathrm{K}$. In the in-plane field configuration,
the traces display qualitatively similar field dependence over the entire
temperature range. In the out-of-plane configuration, an oscillatory
component is clearly resolved at $1.7~\mathrm{K}$ and remains detectable
at $10~\mathrm{K}$, with a reduced amplitude. At $40~\mathrm{K}$ and
above, the oscillations are suppressed below the experimental resolution.

The small sample was measured at approximately ten fixed angular positions between
$0^{\circ}$ and $90^{\circ}$, with an angular step of $10^{\circ}$.
Field sweeps were performed at $1.7$, $3$, $5$, $8$, $12$, $15$, and
$20~\mathrm{K}$. Quantum oscillations remained resolvable up to
$20~\mathrm{K}$.
In addition, an angular sweep was performed to establish the relation
between the rotator position and the crystallographic orientation of the sample.

The cantilever was driven close to its fundamental bending resonance. The unloaded resonance depended on the mounted sample and was $f_{0}^{\mathrm{large}} = 42.550~\mathrm{kHz}$ for the large sample and $f_{0}^{\mathrm{small}} = 44.046~\mathrm{kHz}$ for the small sample.

The fundamental bending mode defines an axis $\mathbf{n}$ about which the free end of the cantilever undergoes a small angular oscillation $\delta\theta$. The external magnetic field $\mathbf{B}$ was rotated in the plane perpendicular to $\mathbf{n}$, such that the static field angle $\theta$ and the oscillatory angular displacement $\delta\theta$ lay in the same plane.

For sufficiently small angular oscillations, the shift of the cantilever's resonance frequency is sensitive to the angular curvature of the magnetic free energy,
\begin{equation}
    k(\theta,B,T)
    =
    \frac{\partial^{2}F}{\partial\theta^{2}},
\end{equation}
where $k$ is the magnetotropic susceptibility. The sample-induced resonance-frequency shift is proportional to $k$.

A Zurich Instruments MFLI lock-in amplifier was used to excite the cantilever and track its resonance frequency using the phase-locked-loop mode. This enabled continuous and sensitive measurement of the resonance-frequency shift as a function of magnetic field, temperature, and field orientation.

\begin{figure}[tbp]
    \centering
    \includegraphics[width=0.55\textwidth]{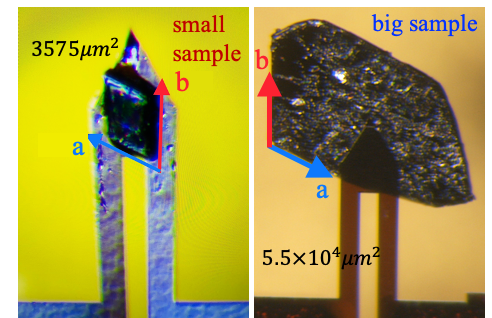}
    \caption{
{\tss    Optical image of the large (right) and small (left) PdCoO$_2$ samples mounted at the tip of the cantilever.
    The crystallographic $a$- and $b$-axes, indicated by the arrows, were determined by EBSD before mounting.}
    }
    \label{fig:pdco2_samples}
\end{figure}

\subsection*{PdCoO$_2$ quantum-oscillation data analysis}

The background-subtracted signal was interpolated onto a uniformly
spaced $1/B$ grid before calculating the fast Fourier transform (FFT).

The frequency resolution associated with an inverse-field interval
$\Delta(1/B)$ is approximately $\delta F \simeq \frac{1}{\Delta(1/B)}$.
The robustness of the extracted spectral features was checked by varying
the field interval and the background-subtraction parameters.

The quantum-oscillation frequencies previously reported for
PdCoO$_2$ are close to $30~\mathrm{kT}$ \cite{Hicks2012}. The corresponding
period in inverse magnetic field is
\begin{equation}
    \Delta\left(\frac{1}{B}\right)
    =
    \frac{1}{F}
    \simeq
    3.3\times10^{-5}~\mathrm{T}^{-1}.
\end{equation}
 In the present measurements, the spacing of
the acquired data points in $1/B$ is comparable to this value. Therefore,
the high-frequency oscillations near $30~\mathrm{kT}$ are undersampled and
cannot be extracted reliably from a direct FFT of the measured signal.

Instead, both samples exhibit a reproducible low-frequency modulation.
The small sample shows dominant spectral weight near
$0.76$--$0.88~\mathrm{kT}$ and a weaker feature near $1.67~\mathrm{kT}$,
while the large sample shows corresponding features near
$0.82$--$0.88~\mathrm{kT}$ and $1.70~\mathrm{kT}$. These low-frequency
features are interpreted as a beat-envelope signal arising from two
unresolved, closely spaced high-frequency oscillations.
Although
the oscillation amplitudes differ substantially between the two samples,
the characteristic frequencies are consistent, indicating that the same
oscillatory response is present in both measurements.

If the oscillatory signal contains two nearby frequencies $F_1$ and $F_2$,
then
\begin{equation}
    \cos\left(2\pi F_1 x\right)
    +
    \cos\left(2\pi F_2 x\right)
    =
    2\cos\left(\pi\Delta F x\right)
    \cos\left(2\pi F_{\mathrm{avg}}x\right),
\end{equation}
where
\begin{equation}
    x=\frac{1}{B},
    \qquad
    \Delta F = \left|F_1-F_2\right|,
    \qquad
    F_{\mathrm{avg}}=\frac{F_1+F_2}{2}.
\end{equation}
Thus, even when the rapid oscillation at $F_{\mathrm{avg}}$ is not
adequately sampled, its interference can produce a slower envelope at a
frequency determined by the separation of the two branches.

Previous de Haas--van Alphen measurements reported two closely spaced
PdCoO$_2$ frequencies near $29.3$ and $30.7~\mathrm{kT}$
\cite{Hicks2012}. Their separation is approximately
$1.4~\mathrm{kT}$. The features observed here near
$0.8$--$0.9~\mathrm{kT}$, together with the weaker structure near
$1.7~\mathrm{kT}$, are therefore most naturally interpreted as a
beat-envelope response and an associated harmonic-like contribution,
rather than as independent fundamental cyclotron orbits.
}

\subsection*{Fermion quantum Monte Carlo for frustrated spin-Peierls systems}

In this section we introduce our AF-QMC formulation of a generic spin-Peierls Hamiltonian. We consider \(S = 1/2\) spins on a lattice interacting with optical phonons. The generic spin-Peierls Hamiltonian reads:
\begin{equation} \label{eq:spin_peierls_hamiltonian}
	\hat{H}  = \sum_{b=\langle i, j \rangle} \left(1 + g \hat{Q}_{b} \right) \left(J_{b}  \vec{\hat{S}}_{i}  \cdot \vec{\hat{S}}_{j}+\sum_{\alpha, \beta}\Gamma_{b}^{\alpha,\beta} \hat{S}_{i}^{\alpha}  \hat{S}_{j}^{\beta} \right) +\sum_{b=\langle i, j \rangle} \left( \frac{\hat{P}_{b}^2}{2 m} + \frac{k}{2} \hat{Q}_{b}^2 \right)
\end{equation}
with \(\alpha, \beta = x,y,z\) and 
\begin{equation} \label{Eq:SN-cr}
	[\hat{S}_{i}^{\alpha}, \hat{S}_{j}^{\beta}]=i \epsilon_{\alpha,\beta,\gamma}\delta_{i,j}\hat{S}_{i}^{\gamma}, \quad [\hat{Q}_{b}, \hat{P}_{b'}]=i \delta_{b,b'}.
\end{equation}
Here the spin-$1/2$ operators $\vec{\hat{S}}_{i}$ reside on a lattice with sites labelled by $i,j$, and the sum runs over nearest-neighbor bonds $b=\langle i,j \rangle$. The phonons with momenta $\hat{P}_{b}$ and displacements $\hat{Q}_b$ reside on each bond. A canonical transformation \(\hat{Q}_{b} \to \hat{Q}_b/g\) and \( \hat{P}_{b} \to g \hat{P}_b\) allows us to set \(g\) to unity by redefining the mass $m$ and spring constant $k$. With this choice, the spin-phonon coupling $\lambda$ and phonon frequency $\omega_0$ read
\begin{equation}
	\lambda = \frac{1}{2k}, \quad \omega_0 = \sqrt{\frac{k}{m}}.
\end{equation}
The spin-phonon interaction is mediated via the $\Gamma_{b}^{\alpha,\beta}$ and $J_{b}$ exchange couplings. While $\Gamma_{b}^{\alpha,\beta}$ defines the potentially frustrated spin model, $J_{b}$ accounts for non-frustrating nearest-neighbor Heisenberg couplings. The special case \(J_b = J\) and \(\Gamma_b^{\alpha,\beta} = 0\) corresponds to the Heisenberg model. The XXZ model is defined by \(J_b = J\) and \(\Gamma_b^{\alpha,\beta} = \delta_{\alpha,\beta} \delta_{\alpha, z} J^z\). Setting \(J_b = J = A \cos(\phi)\) and \(\Gamma_b^{\alpha,\beta} = 2 K \delta_{\alpha, \beta} \delta_{\beta, b} = 2 A \sin(\phi)\) defines the Kitaev-Heisenberg model, with \(A = \sqrt{J^2 + K^2}\). 
{\tss The term $\delta_{\beta,b}$ locks the direction of the spin interaction to the bond direction, thereby defining the bond-dependent spin interactions of the Kitaev model.}
Note that this definition is only valid in lattices with coordination number 3. For the remainder of this work, the Kitaev-Heisenberg model is defined on the honeycomb lattice. The definition of the model for \(\alpha\)-RuCl\(_3\) is given in Eq.~\eqref{Eq:MMKM}.

Our AF-QMC formulation \cite{SatoT21,Inacio25} is based on the Abrikosov fermion representation of the spin operator 
\begin{equation} \label{Eq:AFrep1}
	\vec{\hat{S}}_{i} = \frac{1}{2} \ve{\hat{ f}}^{\dagger}_{i}   \ve{\sigma}  \vec{ \hat{f}} ^{\phantom\dagger}_{i},
\end{equation}
where $\vec{\hat{f}}^{\dagger}_{i}  \equiv  \left(\hat {f}^{\dagger}_{i,\uparrow}, \hat f^{\dagger}_{i,\downarrow} \right) $ is a two-component fermion on site $i$ subjected to the local constraint 
\begin{equation} \label{Eq:AFrep2}
	\hat{n}_{i} = \vec{\hat{f}}^{\dagger}_{i}  \vec{\hat{f}}^{\phantom\dagger}_{i} = 1.
\end{equation}
Here $\ve{\sigma}$ denotes the vector of Pauli spin-$1/2$ matrices. The constraint in Eq.~(\ref{Eq:AFrep2}) is implemented exactly in our fermionic formulation by including a Hubbard-$U$ term on each site and taking the limit $U \rightarrow \infty$. In this representation, the Hamiltonian (\ref{eq:spin_peierls_hamiltonian}) that we simulate reads,
\begin{multline} \label{eq:ham_qmc}
	\hat{H}_{{\rm QMC}}   = \sum_{b} \left( 1 + \hat{Q}_{b} \right) \Biggl[ J_b \left( \frac{1}{4} - \frac{1}{4} \left(\hat{K}_b\right)^2 \right) +  \left( \sum_{\alpha, \beta} \frac{|\Gamma_{b}^{\alpha,\beta} |s^{\alpha\beta}_b }{2} \left(\hat{K}_{b}^{\alpha, \beta}\right)^2 - C_b \right) \Biggr] \\
	+ \sum_b \left( \frac{\hat{P}_{b}^2}{2m} + \frac{k}{2} \hat{Q}_{b}^2 \right) + \frac{U}{2}  \sum_{i} \left(\hat{n}_{i} - 1 \right)^2
\end{multline}
where 
\begin{equation} \label{eq:gauge}
	\hat{K}_b = \sum_\sigma \hat{f}_{i,\sigma}  \hat{f}_{j,\sigma} + \text{h.c.}, \quad \hat{K}_{b}^{\alpha, \beta} = \hat{S}_{i}^{\alpha}  + \frac{\Gamma_{b}^{\alpha,\beta}s^{\alpha\beta}_b} {|\Gamma_{b}^{\alpha,\beta} |}   \hat{S}_{j}^{\beta} , \quad C_{b}=\sum_{\alpha, \beta }\frac{|\Gamma_{b}^{\alpha,\beta} |s^{\alpha\beta}_b}{4}.
\end{equation}
Here \( s^{\alpha\beta}_b = \pm 1 \)~\cite{SatoT21} is a gauge choice in our formulation. In this framework, the operator \(\left(\hat{n}_{i} - 1 \right)^2\) commutes with the Hamiltonian $\hat H_{\rm{QMC}}$, indicating that the $\vec{\hat{f}}$-fermion parity $(-1)^{\hat{n}_{i}}$ is locally conserved. By virtue of this symmetry, introducing a positive Hubbard-$U$ term effectively enforces a projection onto the odd-parity sector where $(-1)^{\hat{n}_{i}} = -1$, dynamically imposing the constraint in Eq.~(\ref{Eq:AFrep2}). Within this sector, the restricted Hamiltonian is $\left. \hat{H}_{\rm{QMC}} \right|_{(-1)^{\hat{n}_{i}} = -1} = \hat{H} + C$, where $C$ is a constant.

Our AF-QMC method was first presented in Ref.~\cite{SatoT21} and later generalized to include phonons in Ref.~\cite{Inacio25}. In the absence of frustrating interactions---such as in the Heisenberg or XXZ models---our AF-QMC formulation does not suffer from the negative sign problem. For the Kitaev-Heisenberg and \(\alpha\)-RuCl\(_3\) models, the sign problem is mild enough to allow simulations of systems with \(N_s = 32\) sites at temperatures \(\beta A \approx 2\)~\cite{SatoT21,SatoT23a}. It was shown in Ref.~\cite{Inacio25} that including phonons does not worsen the severity of the sign problem, so the same system sizes and temperatures remain accessible. We use the Algorithms for Lattice Fermions (ALF) package~\cite{ALF_v2.4} for our AF-QMC simulations.

\subsection*{Sign-problem optimization with machine learning}

Our AF-QMC formulation of the spin-Peierls Hamiltonian \eqref{eq:ham_qmc} offers an extra gauge degree of freedom \(s^{\alpha\beta}_b = \pm 1\) per bond. Although it does not change the physical properties of the model, different choices of \(\{s^{\alpha\beta}_b\}\) modify the severity of the sign problem. The sign is defined by
\begin{equation} \label{eq:sign}
  \text{sgn}(\{s^{\alpha\beta}_b\}) = \left< \frac{\text{Re} (e^{-S(\{s^{\alpha\beta}_b\}, C)})}{\left|\text{Re} (e^{-S(\{s^{\alpha\beta}_b\}, C)}) \right|} \right>_C,
\end{equation}
where \(S(\{s^{\alpha\beta}_b\}, C)\) is the action of configuration \(C\) with gauge configuration \(\{s^{\alpha\beta}_b\}\). The average is taken over the configurations \(C\). We refer the reader to Ref.~\cite{ALF_v2.4} for further details on the calculation and definition of the sign. In general, \(\text{sgn} \in [0, 1]\), and \(\text{sgn} = 1\) corresponds to a sign-problem-free action. In our previous works~\cite{SatoT21,Inacio25}, we only considered translationally invariant gauge configurations \(\{s^{\alpha\beta}\}\). For the Kitaev-Heisenberg model there are only 8 possible gauge configurations, comprising all permutations of \(\{s^{\alpha} = \pm 1,\ \alpha=x,y,z\}\), and we selected the one with the highest sign. 
At high temperatures, \(\beta A \leq 1\), we find that \(\{-1, -1, -1\}\) is optimal, while at low temperatures, \(\beta A \geq 1\), the configuration \(\{-1, 1, 1\}\) (and its two other permutations) yields the highest sign~\cite{SatoT21}. We now ask whether considering a bond-dependent gauge \(\{s^{\alpha}_{i} = \pm 1,\ \alpha=x,y,z,\ i=1,\hdots,N\}\) (where \(N\) is the number of unit cells) can further improve the sign. In principle there is no reason why it should not, since the translationally invariant configurations explored previously represent only a small subset of all possible gauge configurations. Our objective is to find the optimal gauge configuration that maximizes the sign for a given set of model parameters (exchange couplings, temperature, and system size). This is an optimization problem over a parameter space of \(2^{3N}\) configurations. 
 An exhaustive search is clearly infeasible for the system sizes of interest. Instead, we train a neural network (NN) to learn the function \(\text{sgn}(\{s^{\alpha}_{i}\})\) using signs computed by AF-QMC for a set of gauge configurations \(\{s^{\alpha}_{i}\}\), and then use simulated annealing to find the configuration that maximizes the sign predicted by the NN.

We build our model using a feed-forward NN with a variable number of layers and neurons, and use the sigmoid activation function to ensure that the output lies in $[0,1]$. We also use the Huber loss function to weight our model. For the implementation, we use the PyTorch library~\cite{paszke2019pytorchimperativestylehighperformance}. As a first test, we apply our model to a lattice with \(N = 4\) unit cells for the pure Kitaev model at inverse temperature \(\beta A = 4\). The total number of gauge configurations is \(N_c = 2^{3N} = 4096\), allowing us to compare our model against an exhaustive search over all configurations. We train our model on a small subset of approximately \(N_c/2\) configurations. The results are summarized in Tables~\ref{tab:1} and \ref{tab:2}. First, we find that the NN correctly identifies the translationally invariant configurations with maximal sign (Table~\ref{tab:1}), although in some cases it cannot predict the sign value to within \(2\sigma\). Next, we attempt to predict the configuration with the globally maximal sign (Table~\ref{tab:2}). Again, the NN cannot accurately predict the sign value, but it correctly identifies that the highest-sign configurations are those with a local gauge belonging to \(\{-1,-1,-1\}\), \(\{-1, 1, 1\}\), and their permutations---that is, for each bond $b$ the local gauge \(\{s_b^\alpha,\ \alpha=x,y,z\}\) is drawn from all permutations of \(\{-1, 1, 1\}\) or equals \(\{-1,-1,-1\}\). We can therefore conclude that the translationally invariant gauge configurations already lie in the region of configuration space where the sign is maximal, and that the sign does not improve significantly away from these configurations.

As a next step, we tested our model for a larger lattice with \(N = 9\) unit cells at \(\beta A = 2\). The total number of gauge configurations is \(N_c \approx 10^{8}\), precluding an exhaustive search. We train our model on approximately \(10^5\) configurations, corresponding to \(\approx 0.1\%\) of \(N_c\). Again, the model identifies that the optimal gauge configurations have the same local gauge as found for \(N = 4\). For the translationally invariant configurations, the best sign is \(\text{sgn}(\{1,-1,1\}) = 0.606(7)\), while the NN predicts the maximum sign for
\begin{multline}
  G = \{-1,  -1, -1, -1, -1,  1, -1, -1, -1, -1,  1, -1, -1, -1, \\
   -1,  1, -1, -1, -1,  1, -1,  1, -1, -1, -1, -1,1\},
\end{multline} 
with \(\text{sgn}(G) = 0.635(7)\). The NN is thus able to correctly identify the region of configuration space in which the sign is maximal. Within this region, however, distinct gauge configurations yield statistically indistinguishable results to within \(2\sigma\). We therefore conclude that translationally invariant configurations are near-optimal.

\begin{table}[t]
  \centering
  \begin{minipage}{0.45\textwidth}
    \centering
    \begin{tabular}{c c c}
      \hline
      Configuration & AF-QMC & NN \\
      \hline
      \{1,1,1\}         & 0.003(2) & 0.00770 \\
      \{1,1,-1\}        & 0.108(5) & 0.11630 \\
      \{1,-1,1\}        & 0.114(5) & 0.12125 \\
      \{-1,1,1\}        & 0.103(4) & 0.10786 \\
      \{1,-1,-1\}       & 0.305(8) & 0.34955 \\
      \{-1,1,-1\}       & 0.303(6) & 0.30275 \\
      \{-1,-1,1\}       & 0.289(8) & 0.35477 \\
      \{-1,-1,-1\}      & 0.163(5) & 0.20623 \\
      \hline
    \end{tabular}
    \caption{Sign from AF-QMC and NN for all possible translationally invariant gauge configurations for \(N = 4\) and \(\beta A = 4\).}
    \label{tab:1}
  \end{minipage}
  \hfill
  \begin{minipage}{0.53\textwidth}
    \centering
    \begin{tabular}{c c c}
      \hline
      Configuration & AF-QMC & NN \\
      \hline
      \{-1,1,-1,-1,-1,1,1,-1,-1,-1,-1,-1\} & 0.327(8) & 0.32252 \\
      \{1,-1,-1,-1,-1,-1,-1,1,-1,-1,-1,1\} & 0.328(9) & 0.28542 \\
      \{-1,1,-1,1,-1,-1,1,-1,-1,-1,-1,-1\} & 0.329(9) & 0.31312 \\
      \{-1,-1,-1,-1,1,-1,-1,1,-1,-1,1,-1\} & 0.329(10) & 0.29085 \\
      \{-1,-1,1,-1,-1,-1,-1,-1,1,-1,1,-1\} & 0.331(7) & 0.30983 \\
      \{-1,-1,-1,1,-1,-1,1,-1,-1,-1,1,-1\} & 0.331(8) & 0.31240 \\
      \{1,-1,-1,-1,1,-1,-1,-1,1,-1,-1,-1\} & 0.332(8) & 0.30105 \\
      \{1,-1,-1,1,-1,-1,1,-1,-1,-1,-1,-1\} & 0.332(6) & 0.33121 \\
      \{-1,1,-1,1,-1,-1,-1,-1,-1,-1,-1,1\} & 0.336(6) & 0.30416 \\
      \{1,-1,-1,-1,1,-1,-1,1,-1,-1,-1,-1\} & 0.339(7) & 0.27350 \\
      \hline
    \end{tabular}
    \caption{Gauge configurations with maximal sign from AF-QMC and NN prediction for \(N = 4\) and \(\beta A = 4\).}
    \label{tab:2}
  \end{minipage}
\end{table}

 {\tss
\section*{Data availability}
The datasets generated and/or analyzed during the current study are not publicly available due to their size and heterogeneous simulation and experimental data formats, but are available from the corresponding authors on reasonable request.}

{\tss
\section*{Code availability}
The ALF implementation of the spin model used in this work is available on the
\href{https://alf.physik.uni-wuerzburg.de/Hamiltonians/hamiltonians/kit-heis/readme/}{ALF website}.}

\section*{Acknowledgements}
{\tss We are deeply indebted to Arkady Shekhter, whose pivotal contributions have significantly advanced our understanding of RTM experiments.}
The authors thank Brad Ramshaw and Jeroen van den Brink for fruitful discussions.

{\tss
 \section*{Funding}
Computational resources were provided by the Gauss Centre for Supercomputing e.V. through computing time on the GCS Supercomputer SuperMUC-NG at the Leibniz Supercomputing Centre under project number pn73xu. 
Additional scientific support and HPC resources were provided by the Erlangen National High Performance Computing Center (NHR@FAU) of the Friedrich-Alexander-Universit\"at Erlangen-N\"urnberg under the NHR project b133ae.
NHR funding is provided by federal and Bavarian state authorities. 
NHR@FAU hardware is partially funded by the German Research Foundation (DFG) under project number 440719683.
T.S. acknowledges financial support from the Deutsche Forschungsgemeinschaft under grant number SA 3986/1-1 and from the W\"urzburg-Dresden Cluster of Excellence on Complexity, Topology and Dynamics in Quantum Matter ctd.qmat (EXC 2147, project-id 390858490).
F.F.A. and J.I. acknowledge financial support from the German Research Foundation (DFG) under grants AS 120/16-1 and AS 120/16-2  (Project number 493886309) that is part of the collaborative research project SFB Q-M\&S funded by the Austrian Science Fund (FWF) F 86.
K.A.M. acknowledges financial support from the  Austrian Science Fund,  SFB F 86, Q-M\&S. 
S.P. acknowledges financial support from the Austrian Science Fund (SFB F 86, Q-M\&S; Cluster of Excellence quantA, 10.55776/COE1). 
K.A.M., G.R., S.S., V.Z., and H.N. acknowledge financial support from the European Research Council Starting Grant 101078696-TROPIC.
The funders and resource providers had no role in study design, data collection and analysis, decision to publish, or preparation of the manuscript.
}

{\tss
\section*{Author contributions}
J.I., J.S., F.F.A. and T.S. contributed to the theoretical and numerical aspects of the work. G.R., S.S., V.Z., H.N., S.P. and K.A.M. contributed to the experimental aspects and broader materials context of the work. S.K. grew and provided the PdCoO$_2$ single crystals used in this work. All authors participated in the interpretation of the results and provided input on the manuscript.
}

{\tss
\section*{Competing interests}
Fakher F. Assaad is an Editorial Board Member of npj Quantum Materials.
Fakher F. Assaad was not involved in the journal's review of, or decisions related to, this manuscript. 
The authors declare no other competing financial or non-financial interests.
}


\clearpage
\section*{Supplementary Information for: Dynamical magnetotropic susceptibility as a new probe of Kitaev materials and beyond}

\subsection*{Supplementary Note 1: Quantum Monte Carlo results for the XXZ model}
\label{QMC}

\begin{figure}[t]
\centering
\includegraphics[width=0.7\linewidth]{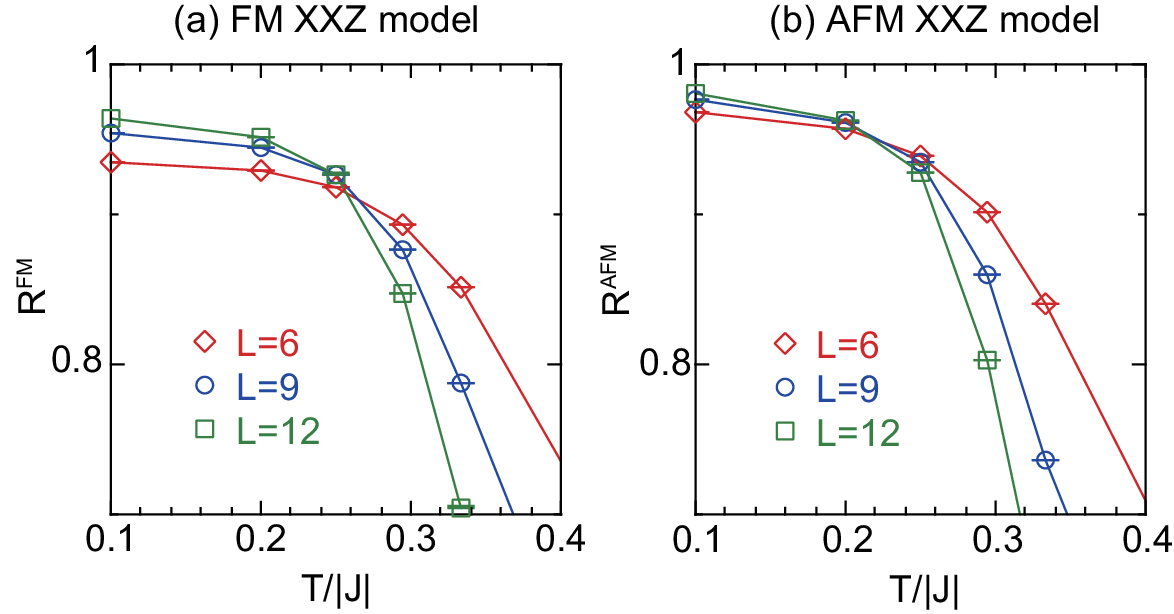}
\caption{\label{fig:CR-SI}
Correlation ratio $R$ for the easy-plane ferromagnetic (FM) order [(a), $(J, J_z) = (-1,0.5)$] and the easy-plane antiferromagnetic (AFM) order [(b), $(J, J_z) = (1,-0.5)$] for different lattice sizes $L$.
As the temperature $T$ decreases, $R$ increases with increasing $L$, indicating the development of long-range order.
At higher temperatures, $R$ decreases with $L$, consistent with a paramagnetic phase.
}
\end{figure}

In the main text, we consider the XXZ model on the honeycomb lattice.
Related XXZ models on the honeycomb lattice have been studied previously using spin-wave theory and series expansions around the Ising limit
~\cite{Weihong1991XXZ,Oitmaa1992Honeycomb}.
The model considered here is defined by
\begin{equation}
  \hat{H}_{\text{Spin}}=\sum_{\langle \ve{i}, \ve{j} \rangle} \left[ J ( \hat{S}^x_{\ve{i}}  \hat{S}^x_{\ve{j}} + \hat{S}^y_{\ve{i}}  \hat{S}^y_{\ve{j}} ) + ( J + J_{z} ) \hat{S}_{\ve{i}}^{z} \hat{S}_{\ve{j}}^{z} \right] - \mu_B\ve{B} \cdot\sum_{\ve{i}}  \hat{g} \hat{\ve{S}}_{\ve{i}},
\end{equation}
and we present QMC results for $J_z/J = -0.5$ and $\mu_B B/|J|=0.1$.

For this choice of parameters, the model favors easy-plane magnetic ordering in the spin-$xy$ direction at low temperatures.

\begin{figure}
\centering
\includegraphics[width=0.8\linewidth]{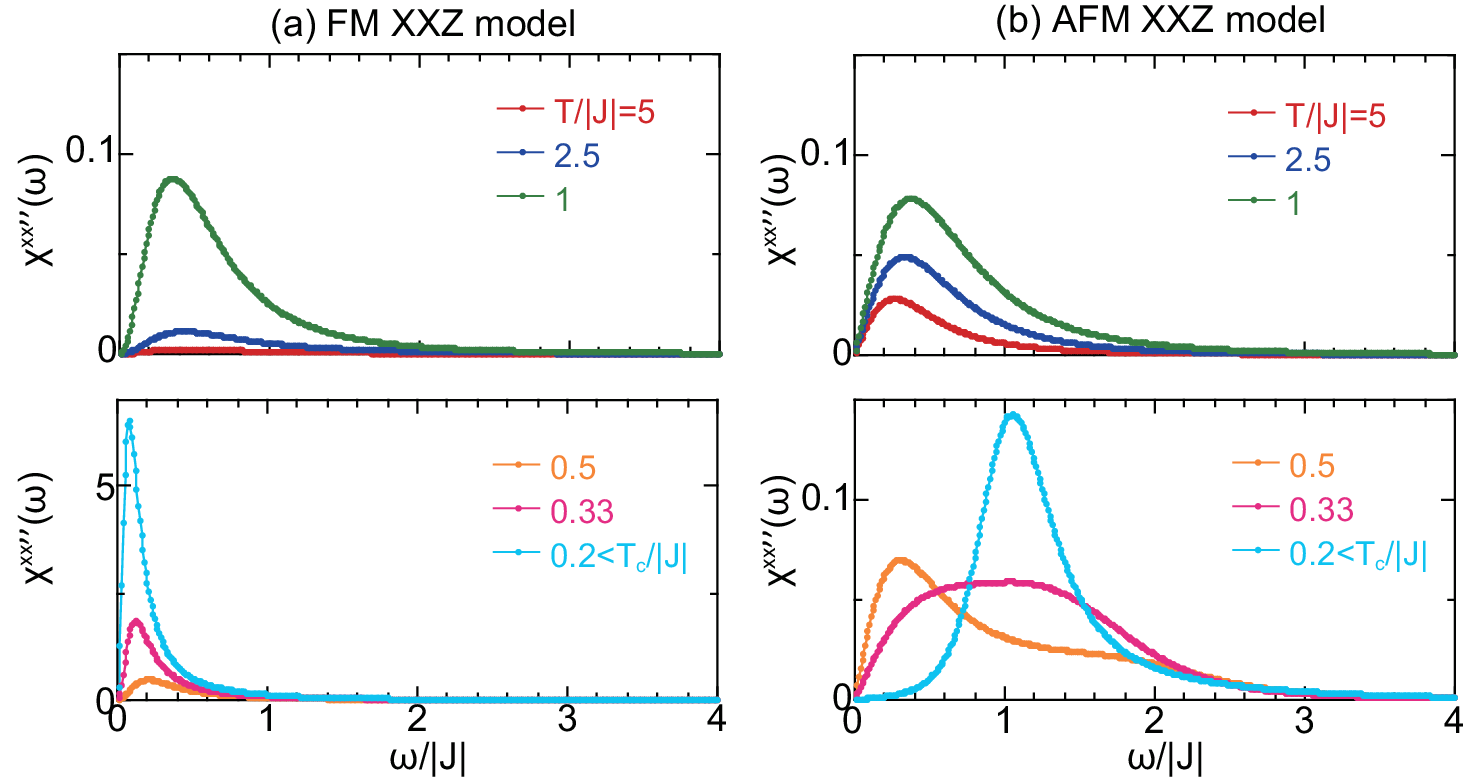}
\caption{\label{fig:chi-XXZ-SI} 
Dynamical spin susceptibility $\chi^{xx}{}''(\omega)$ for the non-frustrated XXZ model on the honeycomb lattice with $\mu_B B/|J|=0.1$. 
(a) Ferromagnetic case $(J,J_z)=(-1,0.5)$ and (b) antiferromagnetic case $(J,J_z)=(1,-0.5)$. 
The magnetic field is applied along $\mathbf{e}_c$ ([001] in the cubic spin basis) with $\mathbf{e}\parallel \mathbf{e}_b$ ([010]). 
For a diagonal $g$ tensor $(g_a,g_b,g_c)=(2,2,1)$, $k''(\omega)$ probes $\chi^{xx}{}''(\omega)$ up to an overall prefactor [Eq.~17 of the main text], such that the temperature evolution is consistent with that shown in Fig.~5 of the main text.
}
\end{figure}

To determine magnetic ordering, we compute the equal-time structure factors
\begin{equation}
C^{\alpha}(\mathbf q)\equiv\frac{1}{L^2}\sum_{\mathbf r, \mathbf r'}\langle  \hat{O}^{\alpha}_{\mathbf r} \hat{O}_{\mathbf r'}^{\alpha}\rangle e^{i \mathbf q \cdot (\mathbf r-\mathbf r')},
\label{eq:C}
\end{equation}
where the antiferromagnetic (AFM) and ferromagnetic (FM) operators are defined as
$  \hat{O}_{\mathbf r}^{\rm AFM} = \hat{S}^{x}_{\mathbf{ r},A} - \hat{S}^{x}_{\mathbf{r},B} $ and $  \hat{O}_{\mathbf r}^{\rm FM} = \hat{S}^{x}_{\mathbf{ r},A} + \hat{S}^{x}_{\mathbf{r},B} $ with $\hat{S}^{\alpha}_{\mathbf{ r},\gamma}= \frac{1}{2} \hat{c}^{\dagger}_{\mathbf{ r},\gamma} \sigma^{\alpha} \hat{c}^{\phantom \dagger}_{\mathbf{ r},\gamma}$.
Here, $\sigma^{\alpha}$ are the Pauli spin-1/2 matrices, $\ve{r}$ labels the unit cell, and $\gamma=A,B$ the sublattices.
From the structure factors we construct the renormalization-group invariant correlation ratios~\cite{Binder1981,Pujari16},
\begin{equation}
R^{\alpha}=1-\frac{C^{\alpha}({\mathbf  q_0}+\delta {\mathbf   q}) }{C^{\alpha}(\mathbf  q_0)},
\label{eq:R}
\end{equation}
where ${\mathbf  q_0}$ is the ordering wave vector and $\delta {\ve q}$ corresponds to the smallest momentum increment allowed by the finite lattice.
In the presence of long-range order, $C^{\alpha}(\mathbf  q_0)$ diverges with system size, implying $R^{\alpha} \to 1$ as $L\to\infty$.
In the disordered phase $R^{\alpha} \to 0$.
At the critical point, $R^{\alpha}$ becomes scale-invariant, leading to a crossing of curves for different system sizes $L$.

We observe a finite-temperature magnetic phase transition, as shown in Figure~\ref{fig:CR-SI}.
Figure~\ref{fig:CR-SI}(a) displays the temperature dependence at $(J, J_z) = (-1,0.5)$.
The onset of long-range easy-plane FM order is signaled by an increase of $R^{\text{FM}}$ with increasing system size $L$.
Figure~\ref{fig:CR-SI}(b) shows the corresponding results for $(J, J_z) = (1,-0.5)$, indicating the emergence of easy-plane AFM order.
We have also verified that no other magnetic ordering tendencies, including easy-axis order along the spin-$z$ direction, develop for these parameter sets.

In addition to the dynamical magnetotropic susceptibility $k''(\omega)$ shown in the main text, we explicitly compute the corresponding dynamical spin susceptibility component $\chi^{xx}{}''(\omega)$ for the same parameter sets and field geometry. 
For $\mathbf{B}\parallel \mathbf{e}_c$ and $\mathbf{e}\parallel \mathbf{e}_b$ in the cubic spin basis, and with a diagonal $g$ tensor, one obtains $g^T(\mathbf{e}\times\mathbf{B})\parallel \mathbf{e}_a$, such that $k''(\omega)$ probes the transverse response $\chi^{xx}{}''(\omega)$ up to an overall prefactor [Eq.~17 of the main text]. 
The resulting $\chi^{xx}{}''(\omega)$, shown in Figure~\ref{fig:chi-XXZ-SI}, exhibits the same temperature evolution as $k''(\omega)$ in both the ferromagnetic and antiferromagnetic cases. 
This confirms that the spectral features discussed in the main text directly reflect the transverse spin dynamics of the XXZ model.

\subsection*{Supplementary Note 2: Hamiltonian in oscillating frame}
The linear response calculations are carried out in the oscillating frame of  the cantilever. Since this is not an inertial  system, it is a valid question to ask if the  spin and orbital inertial terms,  the quantum analogs of
the centrifugal and Coriolis forces,  should be taken into account in our  theory calculation.  The calculation presented here will show  that  we can safely neglect them since they are of the order of  $10^{-8}$ meV.   Since we would like to set the framework at the most fundamental level, we consider the Dirac equation. To estimate  the magnitude of spin and orbital inertial  terms it suffices to consider the non-interacting Dirac equation subject to a vector potential. 
\begin{equation}
 \left[\frac{\hbar}{i} \gamma^\mu \left(\partial_\mu +  \frac{i e}{\hbar} A_\mu(x) \right)  + m c \right] \psi(x) = 0.
\end{equation}
Here  we  use a standard notation \cite{Peskin_book},  with  the $\gamma$-matrices satisfying the Clifford algebra $\{\gamma^\mu, \gamma^\nu\} = 2 g^{\mu\nu}$,  and $g^{\mu\nu}$ the Minkowski metric.  The vector potential $A_\mu(x) = : \left( \frac{\Phi(x)}{c}, - \ve{A}(x) \right) $ and $\partial_\mu =: \left(  \frac{1}{c}\frac{\partial}{\partial  t}, \nabla \right)$.    We  now  consider a  Lorentz  transformation, 
\begin{equation}
\Lambda =   e^{\frac{i}{2} \omega_{\mu,\nu} J^{\mu,\nu}}
\end{equation} 
with generators 
\begin{equation}
    \left(J^{\mu,\nu}\right)^{\alpha}_{\;\; \beta} = i \left( g^{\alpha,\mu} \delta^{\nu}_{\; \; \beta } - g^{\alpha,\nu} \delta^{\mu}_{\;\; \beta } \right)
\end{equation}
and $\omega_{\mu,\nu}$ an antisymmetric tensor. For a rotation about the z-axis with amplitude $\Delta \Theta$ and  frequency $\omega$, the  antisymmetric tensor takes the form 
\begin{equation}
    \omega_{12} = -\omega_{21} = \Delta \Theta \cos(\omega t), \quad \text{and all other components vanish}.
\end{equation}
The Dirac equation in the oscillating frame, $x'^{\mu} =  \Lambda^{\mu}_{\; \; \nu} x^{\nu}$, is obtained by applying the  Lorentz  transformation to the  Dirac equation 
\begin{equation}
   \left[\frac{\hbar}{i} \gamma^\mu \left(\partial^{'}_\mu +  \frac{i e}{\hbar} A{'}_\mu(x') \right)  + m c \right] \psi'(x') = 0. 
\end{equation}
Here, $A{'}_\mu(x') = \Lambda_{\mu}^{\; \; \nu} A_\nu(\Lambda^{-1} x')$,  $\partial^{'}_\mu = \Lambda_{\mu}^{\; \; \nu} \partial_\nu$ and 
\begin{equation}
\Psi'(x') = \Lambda_{\frac{1}{2}} \Psi(\Lambda^{-1} x'), \quad \text{with} \quad \Lambda_{1/2} = e^{\frac{i}{4} \omega_{\mu,\nu} S^{\mu,\nu}} \quad \text{and} \quad S^{\mu,\nu} = \frac{i}{2} [\gamma^\mu, \gamma^\nu]. 
\end{equation}
Taking into  account the  time dependence  of the Lorentz transformation and the fact that it corresponds to a rotation of the spatial components, we  recognize the vector potential of  Eq.~5 of the main text.   In the oscillating frame, it has a time dependence  and hence generates an electric field. The additional terms  that arise from the time dependence of the Lorentz transformation stem from the  time derivative of $\Psi'(x')$.  A  calculation  will  give: 
\begin{equation}
       \frac{i}{\hbar} \frac{\partial }{\partial t } \Psi'(x') =   - \Delta \Theta \sin(\omega t) \omega  \left( \Sigma^{3} + L'^{3} \right) \Psi'(x') + \frac{i}{\hbar} \dot{\Psi}'(x').
\end{equation}
The first term on the right hand side corresponds to the spin and orbital inertial terms, with 
\begin{equation}
    \Sigma^{3}   = \frac{i\hbar}{2} \gamma^1 \gamma^2  = \frac{\hbar}{2} \begin{pmatrix} \sigma^3 & 0 \\ 0 & \sigma^3 \end{pmatrix}, \quad L'^{3} = -i \hbar \left( x'^{1} \partial'_{2} - x'^{2} \partial'_{1} \right),
\end{equation}
the third  component of respectively the spin and  angular momentum  in the  oscillating frame.   Finally,  
\begin{equation}
\dot{\Psi}'(x') = \Lambda_{1/2} \partial_t {\Psi}(\Lambda^{-1} x') 
\end{equation}
is  the time derivative of the spinor taken in the lab frame, and subsequently rotated to the oscillating frame.   At this point, we can carry out the non-relativistic limit of the Dirac equation in the oscillating frame  to obtain the Pauli equation that we  use in the main text but in second quantized form.  

However, even without carrying out the non-relativistic limit, we can estimate the magnitude of the spin and orbital inertial terms. For a cantilever oscillating at a GHz with an oscillation amplitude of $3 \times 10^{-5}$~rad, this energy scale is of the order of $10^{-8}$~meV. Compared to the energy scale of the typical magnetic and electronic Hamiltonians we have in mind, these effects are negligible and can hence  be safely ignored in our calculations.

\end{document}